\documentclass[12pt,a4paper,onecolumn,oneside]{IEEEtran}
\IEEEoverridecommandlockouts
\usepackage{amsfonts,booktabs}
\usepackage{amsmath}
\usepackage{graphicx}
\usepackage{subfigure}
\usepackage{times}
\usepackage{geometry}
\usepackage{setspace}
\let\oldbibliography\thebibliography
\renewcommand{\thebibliography}[1]{%
  \oldbibliography{#1}%
  \setlength{\itemsep}{0pt}%
}

\newenvironment{remark}         {\begin{sideremark}\rm}{\end{sideremark}}
\newtheorem{sideremark}{Remark}

\newtheorem{lemma}{Lemma}
\newtheorem{theorem}{Theorem}

\def\reals{{\rm I\kern-.17em R}}
\def\nats{{\rm I\kern-.17em N}}

\newcommand{\beq}{\begin{equation}}
\newcommand{\eeq}{\end{equation}}
\newcommand{\beqa}{\begin{eqnarray}}
\newcommand{\eeqa}{\end{eqnarray}}
\newcommand{\ole}{\stackrel{\triangle}{=}}

\begin{document}
\title{Spectrum Sharing in Cognitive Radio with Quantized Channel Information}

\author{YuanYuan He and Subhrakanti Dey
\\ Department of Electrical and Electronic Engineering\\ University of Melbourne,
Vic. 3010, Australia\\ e-mail: \{yyhe, s.dey\}@ee.unimelb.edu.au}
\maketitle
\thispagestyle{empty}

\begin{abstract}
%\boldmath 
We consider a wideband spectrum sharing system where a secondary user can share a number of orthogonal frequency bands where  each band is licensed to an individual
primary user. We address the problem of optimum secondary transmit power allocation for its ergodic capacity maximization 
 subject to an average sum (across the bands) transmit power constraint and individual average interference constraints on the 
primary users. The major contribution of our work lies in considering quantized channel state information (CSI)(for the vector channel space consisting of all 
secondary-to-secondary and secondary-to-primary channels) at the secondary transmitter as opposed to the prevalent assumption of full CSI in most existing work. 
 It is assumed that a band manager or a cognitive radio service provider has access to the full CSI information from the secondary and primary receivers and  designs 
 (offline) an optimal power codebook based on the statistical information 
(channel distributions) of the channels and feeds back the index of the codebook to the secondary transmitter for every channel realization in real-time, via a delay-free noiseless limited feedback channel. A modified Generalized Lloyds-type 
algorithm (GLA) is designed for deriving the optimal power codebook, which is proved to be globally convergent and empirically consistent. 
 An approximate 
quantized power allocation (AQPA) algorithm is also presented, that performs very close to its GLA based counterpart for large number of feedback bits 
and is significantly faster. 
We also present an extension of the
 modified GLA based quantized power codebook design algorithm for the case when the feedback channel is noisy. Numerical studies illustrate 
that with only 3-4 bits of feedback, the modified GLA based algorithms provide secondary ergodic capacity very close to that achieved by full CSI and with only as little 
as 4 bits of feedback, AQPA provides a comparable performance, thus making it an attractive choice for practical implementation.

\end{abstract}
 \IEEEpeerreviewmaketitle

\section{Introduction}
\indent Radio spectrum is  a limited and precious natural resource, which, traditionally, is licensed to users by regulatory authorities in a very rigid manner  where in order to avoid interference, the licensed owner has an exclusive right to access the allocated frequency band \cite{Ghasemi07}. Consequently,  as the number of  wireless communication systems and services  grows, the availability of vacant spectrum becomes severely scarce.  However,  recent measurements by the Federal Communications Commission reveal that many portions of spectrum are mostly under utilized or even unoccupied.  This led to the idea of cognitive radio (CR) technology, originally introduced by J. Mitola \cite{Mitola99}, which holds tremendous promise to dramatically improve the efficiency of spectral utilization. The key idea behind CR is that an unlicensed/secondary user (SU) is allowed to communicate over the frequency band originally licensed to a primary user (PU), as long as the transmission of SU does not generate unfavorable impact on the operation of PU.
Effectively, three categories of CR network paradigms have been proposed: interweave, overlay, and underlay \cite{Goldsmith09}.  In the underlay systems, which is the focus of this paper, the SU can transmit even when the PU is present, but the transmitted power of SU should be controlled properly so as to ensure that the resulting interference does not degrade the received signal quality of PU to an undesirable level \cite{Kang09} by imposing the so called interference temperature \cite{Ghasemi07} constraints at PU (average or peak interference power (AIP/PIP) constraint).  This type of CR is also known as the 'spectrum sharing' \cite{Ghasemi07} model. \\
\indent  \cite{Gastpar04} first studied the behavior of capacities of different AWGN channels under received-power constraints (AIP) at the PU receiver (PU-RX),  which showed that for point-to-point non-fading AWGN channels, the capacity performance with  transmit and received power constraints are very similar. The ergodic capacity of narrow band spectrum sharing model with one SU and one or multiple PU under either AIP or PIP constraint at PU-RX in various fading environments was studied by \cite{Ghasemi07}, illustrating that in a fading environment, spectrum access opportunity for the SU significantly increases compared to the AWGN case.  In \cite{Musavian09}, the authors studied optimum power allocation for three different capacity notions under both AIP and PIP constraints.    \cite{Kang09} designed optimal power transmission strategies for maximizing ergodic capacity and outage capacity under various combinations of secondary transmit power constraints and interference  constraints.  %The optimal power control policy of narrow band spectrum sharing model under various combinations of power constraints also has been investigated in \cite{RZhang09} for multiple SUs and multiple PUs in fading cognitive multiple-access channel and
%cognitive broadcast fading channels respectively and \cite{Zhang08} for one SUs and multiple PUs in multiple-access channels. 
\\\indent Most of the above results assume perfect knowledge of full channel state information (CSI) including the SU-TX to PU-RX channels, which is hard to realize in practice. A few recent papers have emerged that address this concern by investigating capacity analysis with imperfect CSI. The effect of imperfect channel estimation in the secondary to primary channels has been investigated in \cite{Musavian092} by considering the channel estimate as a noisy version of the true CSI, and \cite{KRZhang09} proposed a practical design paradigm for cognitive beamforming based on finite-rate cooperative feedback from the PU-RX to the SU-TX. Another recent work \cite{suraweera09}
also considers imperfect CSI for the SU-TX to PU-RX channel in the form of noisy channel estimate and quantized channel information and investigates the effect of such 
imperfect CSI on the capacity performance of the secondary user, while assuming that the SU-Tx has full knowledge of the SU-Tx to SU-Rx channel. Finally, 
\cite{marques_giannakis09} studies the issue of channel quantization for resource allocation via the framework of utility maximization in OFDMA based 
cognitive radio networks, but does not investigate the joint channel partitioning and rate/power codebook design problem.  Indeed, the lack of a rigorous and systematic 
design methodology for quantized resource allocation algorithms in the context of cognitive radio networks forms the key motivation for our work.  In this paper, we investigate an ergodic capacity optimization problem for the secondary user where quantized information about the vector channel space consisting of SU-Tx to SU-Rx channels and SU-Tx to PU-Rx channels is available to the SU-Tx via a limited feedback channel without 
delay. 
 We consider a wideband spectrum sharing system where one SU shares $M$ different frequency bands with $M$ PU's, each PU using a separate 
band. We address the problem of ergodic capacity maximization 
of the secondary user subject to an average sum (across the bands) transmit power constraint on the secondary user and individual average interference constraints on the 
primary users, using quantized channel information. To this end, we assume the availability of an entity called a band manager (or a CR service provider)
who has access to the full CSI including all 
secondary-to-secondary and secondary-to-primary channels. It designs (offline) an optimal power codebook based on the statistical information 
(channel distributions) of the channels and in real-time, feeds back the index of the codebook to the secondary transmitter for every channel realization, via a limited feedback link. The secondary transmitter then uses the corresponding power code vector for its transmission.  \\
\indent We make the following key contributions: (1) 
We first  present, very briefly, a systematic algorithm for optimal power allocation with full channel side information (CSI) at the secondary transmitter. This is a minor extension of the results in \cite{Kang09} to the multiple PU case. However, the novelty lies in exactly characterizing the optimal power allocation 
policy based on the relationship between the available total average SU transmit power and the individual average interference levels at the PU receivers.
(2) Next, we present a modified Generalized Lloyd's type algorithm (GLA) for designing the optimal power codebook
 using quantized channel information.  For easier exposition, we focus on the narrowband case first and present the quantized power allocation algorithm, where we prove 
 that the modified GLA based power codebook design algorithm is globally convergent and empirically consistent. We provide a number of useful and interesting 
 properties of the quantized powers. Then we present a complete description of the optimal power codebook design algorithm for the wideband spectrum sharing case under 
 the average transmit power and average interference constraints. We believe this paper is the first to provide a systematic quantized power allocation algorithm with 
 limited feedback for the spectrum sharing scenario in cognitive radio. (3) Although an offline algorithm, GLA based quantizer designs usually require a large number of training samples and can be computationally expensive. We 
 therefore design an approximate quantized power allocation algorithm based on the derived properties of the power codebook, which is computationally much faster. 
(4) We then generalize the  modified GLA based algorithm for quantized power allocation algorithm to the case where the limited feedback channel is noisy but memoryless.
(5) We present a comprehensive set of numerical results that illustrate (i) how the modified GLA-based power codebook can achieve a secondary ergodic capacity with only 3-4 bits of feedback, that is very close to the capacity with full CSI, (ii) how the performance of the approximate quantized power allocation algorithm is almost indistinguishable from that of the GLA-based algorithm with $B\geq 4$ bits of feedback and (ii) how the performance of the quantized power allocation degrades when 
 the noisy feedback channel error probability increases. 
 
The rest of the paper is organized as follows. Section \ref{model} presents the system model and assumptions about the spectrum sharing problem with limited feedback. 
In Section \ref{fullcsi}, we present the optimal power allocation policy when the secondary transmitter has full CSI and discuss various special cases.
In Section \ref{GLA}, we present the modified GLA based quantized power codebook design algorithms for the narrowband case followed by the wideband case. We present results on global convergence and empirical consistency of the GLA based algorithms and some prove some useful properties of the quantized power code vectors. These 
properties are then used to design an approximate quantized power allocation algorithm suitable for moderate to large number of feedback bits that has a much faster convergence time compared to its GLA counterpart. In Section \ref{noisyfeedback}, we provide a modified 
GLA based power codebook design algorithm for   a noisy limited feedback channel model. Numerical results are presented in Section \ref{sec:simul} and finally, concluding 
remarks and possible extensions are presented in Section \ref{conclu}. All proofs are relegated to the Appendix unless otherwise mentioned.
 
\section{System Model and Problem Formulation} \label{model}
\begin{figure}[h]
\centering
\includegraphics[scale=0.5]{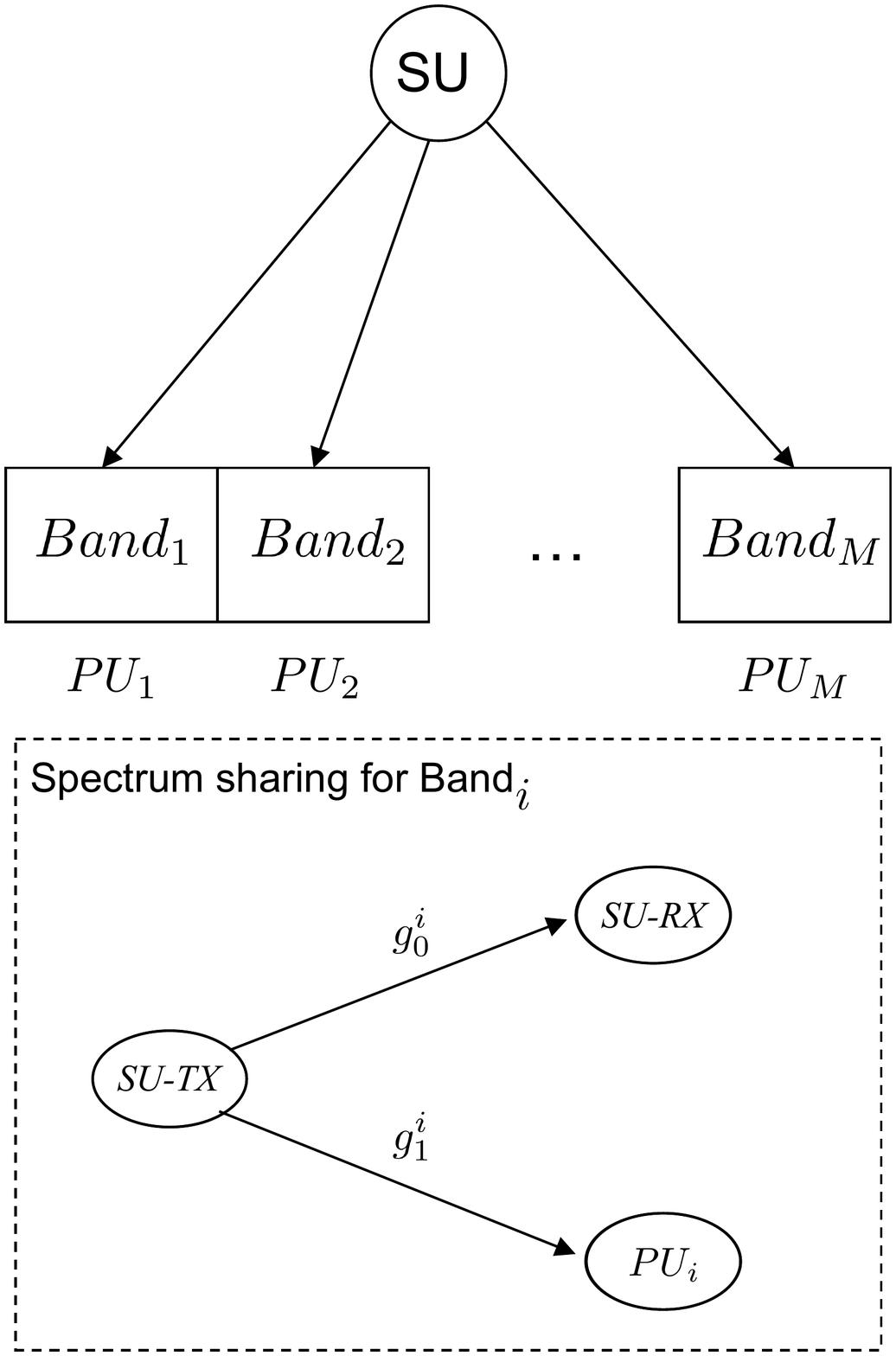}
\caption{System model for wideband spectrum-sharing scenario.}
\label{f1}
\end{figure}

\indent We consider  a wideband spectrum sharing
scenario with one SU and Multiple PUs, as shown in Fig. {\ref{f1}}, where a SU is allowed to use M parallel orthogonal frequency bands ($Band_1$ to $Band_M$) which are 
individually licensed to $PU_1$, $\dots$, $PU_M$ respectively.  Regardless of the on/off status of $PU_i$, SU uses the $i$-th channel  as long as the impact of the secondary transmission does
 not substantially degrade the received signal quality  $PU_i$. It is assumed that the the channels between the secondary transmitter  (SU-TX) and secondary (SU-RX) receiver and those between the secondary transmitter and the each primary receiver are all block fading additive white Gaussian noise (BF-AWGN) channels. 
 Let $g^i_0 \in \reals_{+}$ and $g^i_1 \in \reals_{+}$ denote the real-valued instantaneous channel power gains for the link between the SU-TX and the receiver of 
 $PU_i$ and $i$-th channel between the SU-TX and SU-RX, respectively, where $\reals_{+}$ denotes the set of nonnegative real numbers. These 
channels are assumed to be stationary ergodic with absolutely continuous probability density functions (pdf) $f_0(g^i_0)$ and $f_1 (g^i_1)$. For analytical simplicity, 
the interference from $PU_i$-TX to SU-RX is neglected (similarly as in \cite{Ghasemi07,Kang09}). In the case where the interference caused by the 
primary transmitter at the secondary receiver is significant, the SU ergodic capacity results derived in this paper can be taken as upper bounds on the 
actual capacity under primary-induced interference.    All $g^i_0$ and $g^i_1$  ($i=1,\dots,M$) are statistically 
mutually independent and, without loss of generality ({\em w.l.o.g}), are assumed to have unity mean. Similarly, additive noises for each channel are independent Gaussian random variables with zero mean and unit variance {\em w.l.o.g}.  
When $M=1$, this system becomes a typical narrowband spectrum sharing model considered in \cite{Ghasemi07}\cite{Zhang09}\cite{Kang09}.\\
\indent  Given a channel realization $\textbf {g}_0 \ole \{g^1_0,\dots,g^M_0\} $ and $\textbf {g}_1 \ole 
\{g^1_1,\dots,g^M_1\} $, we assume that a channel side information (CSI) $\eta({\mathbf g_0}, {\mathbf g_1})$ is available 
at the SU-TX. The power allocated at the SU-TX on the M parallel SU links is represented by  the vector $\textbf{p}(\eta(\textbf {g}_0, \textbf {g}_1))=\{p_1(\eta(\textbf {g}_0, \textbf {g}_1)),\ldots,p_M(\eta(\textbf {g}_0, \textbf {g}_1))\}$, the ergodic capacity  of the SU for this wideband spectrum sharing system can be expressed as
\begin{eqnarray}
C=\frac{1}{M}\sum^M_{i=1}E[\log(1+g^i_1p_i(\eta(\textbf {g}_0, \textbf {g}_1)))]
\end{eqnarray} 
where, for simplicity, we have ignored the factor $\frac{1}{2}$ at the front of the capacity expression and $\log$ represents the natural logarithm. 
 A common way to protect PU's received signal quality is by imposing either an average or a peak interference power (AIP/PIP) constraint at PU-RX \cite{Ghasemi07}\cite{Zhang09}\cite{Kang09}, although other forms of PU quality of service constraint such as PU's capacity loss and PU's outage 
 probability \cite{zhang_sigprocmag}. 
 %and these two kinds of constraints can be expressed as
%\begin{eqnarray}
%E[g^i_0 p_i(\textbf {g}_0, \textbf {g}_1)] \leq Q^i_{avg},\\
%g^i_0 p_i(\textbf {g}_0, \textbf {g}_1)\leq Q^i_{peak}
%\end{eqnarray} 
 It was shown in \cite{Zhang09} that an AIP constraint is more favorable than  a peak constraint especially in the context of transmission over fading channels, since the AIP constraint is more flexible and can achieve  larger SU capacity results with less PU capacity loss than those achieved by PIP.\\
\indent Motivated by this observation,  we consider the following optimal power allocation scheme that maximizes the ergodic capacity  of SU in a wideband spectrum sharing scenario, under an AIP constraint at each $PU_i$-RX and an average sum transmit power constraint (ATP) for the SU, given by,
%\begin{eqnarray}
%\frac{1}{M}\sum^M_{i=1}E[p_i(\textbf {g}_0, \textbf {g}_1)] \leq P_{avg}
%\label{aver}
%\end{eqnarray} 
%Formulating this as an optimization problem, we have
\begin{eqnarray}
&&\max_{p_i(\eta(\textbf {g}_0, \textbf {g}_1)) \geq 0,   \forall  i} ~~\frac{1}{M}\sum^M_{i=1}E[\log(1+g^i_1p_i(\eta(\textbf {g}_0, \textbf {g}_1)))]\nonumber\\
&&~~~~~~s. t. ~~~~~~~E[g^i_0 p_i(\eta(\textbf {g}_0, \textbf {g}_1))] \leq Q^i_{avg},    ~~\forall  i, \; 
\frac{1}{M}\sum^M_{i=1}E[p_i(\eta(\textbf {g}_0, \textbf {g}_1))] \leq P_{avg}
\label{Q1}
\end{eqnarray} 
 
In the next section, we present the optimal power allocation results assuming that full channel state information (CSI) is available at the SU-Tx (i.e, 
$\eta(\textbf {g}_0, \textbf {g}_1) = (\textbf {g}_0, \textbf {g}_1)$), followed by the case of quantized channel information (or limited feedback) in Section IV, where 
$\eta(\textbf {g}_0, \textbf {g}_1)$ represents a deterministic index mapping scheme, such that 
$\eta(\textbf {g}_0, \textbf {g}_1) = j, \; j \in \{1,2, \ldots, L\}$ when the instantaneous channel gains $(\textbf {g}_0, \textbf {g}_1)$ belong to a 
carefully constructed partition ${\cal R}_j$ of the channel space $\reals_{+}^M \times \reals_{+}^M$.  

\section{Optimal Power Allocation with Perfect Channel State Information }\label{fullcsi}
\indent In this section, we assume that SU-TX has perfect knowledge of $\textbf {g}_0$ and $\textbf {g}_1$ (full CSI at the transmitter), that is, 
$\eta(\textbf {g}_0, \textbf {g}_1) = (\textbf {g}_0, \textbf {g}_1)$. 
 It is easy to verify that the problem given in (\ref{Q1}) is a convex optimization problem. By applying the necessary and sufficient Karush-Kuhn-Tucker (KKT) conditions for optimality, the optimal power allocation can be easily shown to be
 
 %from the following equations: 
%\begin{eqnarray}
%&&\lambda\geq 0, ~\mu_i \geq 0, ~p_i(\textbf {g}_0, \textbf {g}_1) \geq 0 ~~~\forall i\\
%\label{c}
%&&\lambda(\frac{1}{M}\sum^M_{i=1}E[p_i(\textbf {g}_0, \textbf {g}_1)] - P_{avg})=0\\
%\label{b}
%&&\mu_i (E[g^i_0 p_i(\textbf {g}_0, \textbf {g}_1)] - Q^i_{avg})=0    ~~\forall  i\\
%\label{a}
%&&\frac{\partial L(\textbf{P},\lambda,\textbf{u})}{\partial {p_{i}}}=\frac{g^i_1}{1+g^i_1p_i}-\lambda-\mu_i g^i_0=0,   ~~\forall  i
%\end{eqnarray} 
%and from (\ref{a}), we can obtain
\begin{eqnarray}
p_i^*(\textbf {g}_0, \textbf {g}_1) =(\frac{1}{\lambda^f+\mu_i^f g^i_0}-\frac{1}{g^i_1})^+
\label{d}
\end{eqnarray} 
where  $\lambda^f$ and $\mu_i^f$ are the nonnegative Lagrange multipliers
associated with the ATP constraint and the AIP constraint of $PU_{i}$ respectively, and 
$(x)^{+} = max(x,0)$. 
This solution is clearly a minor extension of the narrowband 
result in \cite{Kang09}. However,  in the wideband case ($M >1$), it should be noted that determining the optimal power allocation scheme 
 involves obtaining the optimal values of the $(M+1)$ Lagrange multipliers. Since  all the constraints in Problem (\ref{Q1}) 
 may not hold with equality simultaneously, it is difficult to determine  $\lambda^f$ and $\mu_i^f,~~\forall i$.   Although they can be obtained by, e.g., the ellipsoid method \cite{RZhang09}, this procedure can be time consuming.  Thus motivated, we present a complete solution to Problem (\ref{Q1}), summarized in the 
 following theorem (here the term ``iff" refers to ``if and only if") 
\indent \begin{theorem}
With perfect channel information $\eta(\textbf {g}_0, \textbf {g}_1) = (\textbf {g}_0, \textbf {g}_1)$ at the SU-TX, the optimal power allocation for problem  (\ref{Q1}) is given by
\begin{eqnarray}
p_i^*(\textbf {g}_0, \textbf {g}_1) = 
\begin{cases}
(\frac{1}{\mu_i^f g^i_0}-\frac{1}{g^i_1})^+   ~~~~~~~~~~~~~~~~~~~~~~~~~~~~~~~~~~~~\text{iff}~~ P_{avg}\geq \frac{1}{M}\sum^M_{i=1}E[(\frac{1}{\mu_i g^i_0}-\frac{1}{g^i_1})^+]\\
%(\frac{1}{\lambda^f}-\frac{1}{g^i_1})^+   ~~~~~~~~~~~~~~~~~~~~~~~~~~~~~~~~~~~~~~~\text{iff}~~~  P_{avg}\leq \frac{1}{M}\sum^M_{i=1} Q^i_{avg} \\
\begin{cases}
(\frac{1}{\lambda^f}-\frac{1}{g^i_1})^+   ~~~~~\text{iff}~  E[(\frac{1}{\lambda^f}-\frac{1}{g^i_1})^+] \leq Q^i_{avg}\nonumber\\
~~~~~~~~~~~~~~~~~~~~~~~~~~~~~~~~~~~~~~~~~~~~~~~~~~~~\text{otherwise}\nonumber\\
(\frac{1}{\lambda^f+\mu_i^f g^i_0}-\frac{1}{g^i_1})^+~~~~~\text{otherwise}
\end{cases}\\
\end {cases}
\label{eq:fullcsioptp}
\end{eqnarray}
\label{t1}
 \end{theorem} 
\indent\indent\indent \textit {Proof:}  See Appendix A for a proof. \\
%The following remark explains the above result.
%\indent \begin{remark}  If $P_{avg}\geq \frac{1}{M}\sum^M_{i=1}E[(\frac{1}{\mu_i g^i_0}-\frac{1}{g^i_1})^+]$,
%the ATP constraint is inactive, but all AIP constraints hold with equality simultaneously.
%%   If $P_{avg}\leq \frac{1}{M}\sum^M_{i=1} Q^i_{avg} $,only ATP constraint holds with equality, and all AIP constraints are inactive. 
%  If $P_{avg}$ does not satisfy the above condition, 
%the ATP constraint must hold with equality, and in this case, if $E[(\frac{1}{\lambda^f}-\frac{1}{g^i_1})^+] \leq Q^i_{avg}~~\forall i$, i.e. all AIP constraints are inactive, otherwise, at least one AIP constraint is active.\label{r3}\end{remark}   
One can also easily obtain the following {\em special cases} (we do not provide the proofs due to space constraints as they are straightforward): 
\begin{enumerate}
\item When $M=1$(narrowband spectrum sharing case), from theorem \ref{t1},  the condition $E[(\frac{1}{\lambda^f}-\frac{1}{g^1_1})^+] \leq Q_{avg})$ becomes $P_{avg}\leq Q_{avg}$ (note that we have removed the superscript from $Q_{avg}$ as there is only one primary user), and the optimal power allocation solution specialises to the one presented in \cite{Kang09}.\\
\item When $\mu_i=0 ~~\forall i$,
\begin{eqnarray}
p_i^*(\textbf {g}_0, \textbf {g}_1)=(\frac{1}{\lambda^f}-\frac{1}{g^i_1})^+ , ~~~~E[(\frac{1}{\lambda^f}-\frac{1}{g^i_1})^+] \leq Q^i_{avg} ~~\forall i 
\end{eqnarray}
where $\lambda^f$ is given by $\frac{1}{M}\sum^M_{i=1}E[(\frac{1}{\lambda^f}-\frac{1}{g^i_1})^+   ] = P_{avg} $. For this case, if  additionally $\textbf {g}_1$ are independent and identically distributed, we can simplify the condition $E[(\frac{1}{\lambda^f}-\frac{1}{g^i_1})^+] \leq Q^i_{avg}~~\forall i$ as $P_{avg}\leq min(Q^1_{avg},\dots,Q^M_{avg})$. \\
\item  If  $Q^1_{avg}=\dots=Q^M_{avg}=Q_{avg}$ and both $\textbf {g}_0$ and $\textbf {g}_1$ are independent and identically distributed, the optimal power allocation policy is to assign equal power to each SU link, which is identical to the power allocation policy for the $M=1$ case.
\end{enumerate}
 Appealing to the convexity of Problem (\ref{Q1}), one can  show that
in Theorem \ref{t1}, one of the cases must hold, and the  corresponding power allocation scheme must be the global optimal solution for the original problem (\ref{Q1}).
An algorithm can be then easily designed to obtain $p_i^*(\textbf {g}_0, \textbf {g}_1)$, and the associated non-zero Lagrange multipliers can be obtained by solving 
the KKT optimality conditions numerically (e.g, via a bisection search).  
 
\section{Optimum Quantized Power Control with Finite-Rate Feedback} \label{GLA}
\indent The assumption of full CSI at the SU-TX (especially that of $\textbf {g}_0$)  is usually unrealistic in  practical systems.  In this section, we are therefore interested in 
designing power allocation schemes based on quantized ($\textbf {g}_0$, $\textbf {g}_1$) information acquired via a no-delay and error-free feedback link with limited rate.  Here we assume that there is an entity (such as  CR service provider or a {\em band manager} \cite{peha09})
 who can obtain  perfect information on $\textbf {g}_1$  from SU-RX or SU base stations and perfect information on $\textbf {g}_0$  from PU base stations, presumably 
 over a wired link,
  and then forward some appropriately quantized CSI to SU-TX (and SU-RX for decoding purposes) through the feedback link.  More specifically,  given B bits of feedback, a power codebook  $\cal {P}$$=\{ \textbf {P}_1, \dots, \textbf{P}_L\}$ (where $\textbf{P}_j = \{p_{1j},\dots,p_{Mj}\},~~j=1,\dots,L$) of cardinality $L=2^B$, is designed off line purely on the basis of the statistics of $\textbf {g}_0$, $\textbf {g}_1$. This codebook is known {\em a priori} by both SU-TX and SU-RX. The vector space of ($\textbf {g}_0$, $\textbf {g}_1$), is thus partitioned into $L$ regions $\cal{R}$$_1, \dots,$$\cal{R}$$_L$ using a quantizer $\cal{Q}$ (codebook element $\textbf{P}_j$ represents the power level used in $\cal{R}$$_j$ ). The CR service provider/band manager  maps the current instantaneous  ($\textbf {g}_0$, $\textbf {g}_1$) information into one of $L$ integer indices and sends the corresponding index to  the SU-TX via the feedback link (e.g., if the current ($\textbf {g}_0$, $\textbf {g}_1$) falls in $\cal{R}$$_j$, then $\eta(\textbf {g}_0$, $\textbf {g}_1) = j$ will be conveyed  back to SU-TX). The SU-TX will use the associated power codebook element (e.g.,~if the feedback signal is $j$, then $\textbf{P}_j$ will be used as the transmission power) to adapt its transmission strategy. \\
\indent Let $Pr ({ \cal {R}}_j)$, $E[\bullet|{\cal{R}}_j]$ denote $Pr ((\textbf {g}_0, \textbf {g}_1)\in { \cal {R}}_j)$ (the probability that $(\textbf {g}_0, \textbf {g}_1)$ falls in the region ${ \cal {R}}_j$) and $E[\bullet|(\textbf {g}_0, \textbf {g}_1)\in {\cal{R}}_j]$, respectively. Then the secondary ergodic capacity maximization problem (\ref{Q1}) with limited feedback can be formulated as
\begin{eqnarray}
\max_{\textbf{P}_j \geq 0,   \forall  j}~  \sum^L_{j=1}(\frac{1}{M}\sum^M_{i=1}E[\log(1+g^{i}_{1}p_{ij})| {\cal{R}}_j]) Pr ({ \cal {R}}_j)\nonumber\\
s. t. ~~\sum^L_{j=1}E[g^{i}_0 p_{ij}| {\cal{R}}_j] Pr ({ \cal {R}}_j)\leq Q^i_{avg},    ~\forall  i, \; 
\sum^L_{j=1}(\frac{1}{M}\sum^M_{i=1}p_{ij}) Pr ({ \cal {R}}_j)\leq P_{avg}
\label{Q5}
\end{eqnarray} 
Our objective is thus the joint optimization of the channel partition regions  and the power codebook such that the ergodic capacity of SU is maximized
under the above average transmit power and average interference constrains.
\subsection{ Narrowband spectrum-sharing case }
\indent For ease of exposition, we first look at the relatively simpler case of  $M=1$ (where SU shares a narrowband spectrum with only one PU). 
For simplicity (with some abuse of notation), 
let  $p_{j}, g_1, g_0, Q_{avg}$ represent $p_{1j}, g^1_1, g^1_0, Q^1_{avg}$ respectively. Thus problem (\ref{Q5}) with $M=1$ becomes,
\begin{eqnarray}
&&\max_{p_j \geq 0,   \forall  j}  ~\sum^L_{j=1}E[\log(1+g_{1}p_{j})| {\cal{R}}_j] Pr ({ \cal {R}}_j)\nonumber\\
&&~~s. t. ~~\sum^L_{j=1}E[g_0 p_{j}| {\cal{R}}_j] Pr ({ \cal {R}}_j)\leq Q_{avg}, \;\sum^L_{j=1}p_{j}Pr ({ \cal {R}}_j)\leq P_{avg}
\label{Q6}
\end{eqnarray} 
We solve the problem (\ref{Q6}) based on the Lagrange duality method. First we write the Lagrangian of above problem as
{\begin{eqnarray}
 L(P,\lambda, \mu) =\sum^L_{j=1} E[\log(1+g_1p_j)-\lambda p_j-\mu g_0 p_j |{\cal{R}}_j]Pr ({ \cal {R}}_j)+\lambda P_{avg}+\mu Q_{avg}
\end{eqnarray}}
where $\lambda$ and $\mu$ are the nonnegative Lagrange multipliers
associated with the ATP constraint and AIP constraint respectively.
The Lagrange dual function $g(\lambda, \mu)$ is defined as 
\begin{eqnarray}
\max_{p_j \geq 0 ~\forall j}  \sum^L_{j=1} E[\log(1+g_1p_j)-\lambda p_j-\mu g_0 p_j|{\cal{R}}_j]Pr ({ \cal {R}}_j)
\label{Q7}
\end{eqnarray} 
and the corresponding dual problem is
$\min_{\lambda\geq 0, ~\mu\geq 0}  g(\lambda, \mu)+\lambda P_{avg}+\mu Q_{avg}$. \\
\indent We first consider solving the optimization problem (\ref {Q7}) with fixed $\lambda$ and $\mu$.
To this end, we employ an algorithm similar to a Generalized Lloyd Algorithm (GLA) \cite{Linde80,Gersho92} to design an optimal codebook for problem (\ref {Q7}), which is based on two optimality conditions : 1) optimum channel partitioning  for a given codebook,  also called the nearest neighbor condition (NNC) in the context of traditional vector quantization (VQ), and  2) optimum codebook design for a given partition,  also known as the centroid condition (CC) (in the context of VQ) \cite{Gersho92}.  GLA is usually initialized with a random choice of codebook, and then the above two conditions are iterated until some pre-specified convergence criterion is met.  The same procedure is used here for designing an optimal quantizer $\cal{Q}$, but the design criterion for our case is minimizing the difference between the capacity with perfect CSI and the capacity with quantized power allocation under the given constraints. This amounts to designing an optimal power codebook $\cal{Q}$ that maximizes the Lagrangian function for quantized CSI,
$ \sum^L_{j=1} E[\log(1+g_1p_j)-\lambda p_j-\mu g_0 p_j|{\cal{R}}_j]Pr ({ \cal {R}}_j)$. We call the corresponding quantized power allocation algorithm for a given 
$\lambda, \mu$ as a {\em modified GLA}.  \\
\indent In practice, this modified GLA is implemented using a sufficiently large number of training samples (channel realizations for $g_0, g_1$). Beginning with a random initial codebook, one can design the optimal partitions using the fact that the optimal partitions satisfy 
 ${\cal R}_j=\{(g_0, g_1): (\log(1+g_1p_j)-\lambda p_j-\mu g_0 p_j)\geq ( \log(1+g_1p_n)-\lambda p_n-\mu g_0 p_n), \forall n \neq j\}$ 
 where ${\cal R}_j$ is the corresponding partition region  for power level $p_j$ in the codebook, and ties are broken arbitrarily.
  Once the optimal partitions are designed, the 
 new optimal power codebook is found by solving for $\text{argmax}_{p_j\geq 0} E[\log(1+g_1p_j)-\lambda p_j-\mu g_0 p_j|{\cal{R}}_j]Pr ({ \cal {R}}_j)$, 
 $\forall j=1,2,\ldots, L$. Given a 
 partition, this optimization problem is convex and by using the KKT conditions, one can obtain the optimal power as $\max(p_j^*, 0)$, where $p_j^*$ is the solution to the equation $E[\frac{g_1}{1+g_1p_j}-(\lambda+\mu g_0)| {\cal{R}}_j]=0$. These two steps are repeated until the resulting ergodic capacity converges within a prespecified accuracy.
One needs to note that GLA cannot in general guarantee global optimality, since the two optimality conditions (NNC and CC) mentioned above are just necessary conditions \cite{Gersho92}.  Thus it is possible that the our resulting quantizer is only  locally optimal. While convergence of our modified GLA follows immediately by 
noting that the Lagrangian $ \sum^L_{j=1} E[\log(1+g_1p_j)-\lambda p_j-\mu g_0 p_j|{\cal{R}}_j]Pr ({ \cal {R}}_j)$ 
is non-decreasing at each iteration and is upper bounded (due to the finite average transmit power and average interference constraints), it is important and 
instructive
to state a more formal result along the lines of \cite{Sabin86}.  Since GLA is initialized with a random codebook and the optimal partitions and codevectors 
are found using training samples drawn from empirical distributions, it is crucial that GLA is globally convergent with respect to the choice of initial codebooks
and empirically consistent. For more formal definitions of these two properties, see \cite{Sabin86}. Under the assumption of absolutely continuous 
fading distributions for $g_0, g_1$ and mild regularity assumptions satisfied by these distributions, one can show that the modified GLA satisfies the conditions 
for global convergence and empirical consistency stated in \cite{Sabin86} and thus we have the following result: 
\indent \begin{theorem}
The modified GLA that solves the optimization problem  (\ref{Q7}) satisfies the global convergence and empirical consistency properties of \cite{Sabin86}.
\label{the2}
\end{theorem}
\indent \indent \indent \textit{Proof:} See Appendix A for a proof of this result.\\

\begin{figure}
%\begin{minipage}[t]{0.45\linewidth}
\centering
\includegraphics[scale=0.9]{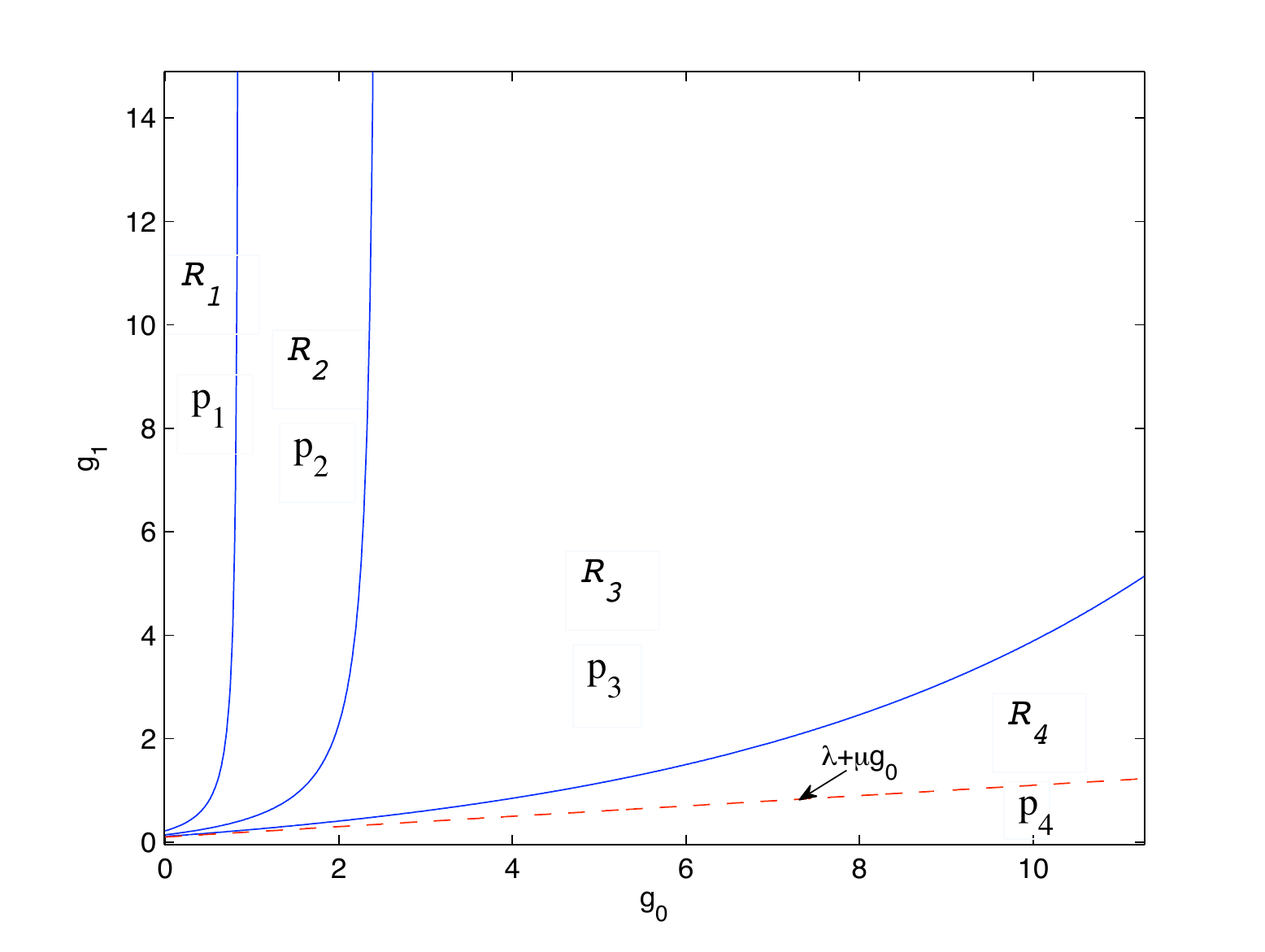}
\caption{The structure of optimum partition regions with $B=2$ bits feedback given $\lambda=\mu=0.1$}
\label{f2}
 \end{figure}

\begin{figure}[h]
\centering
\includegraphics[scale=0.9]{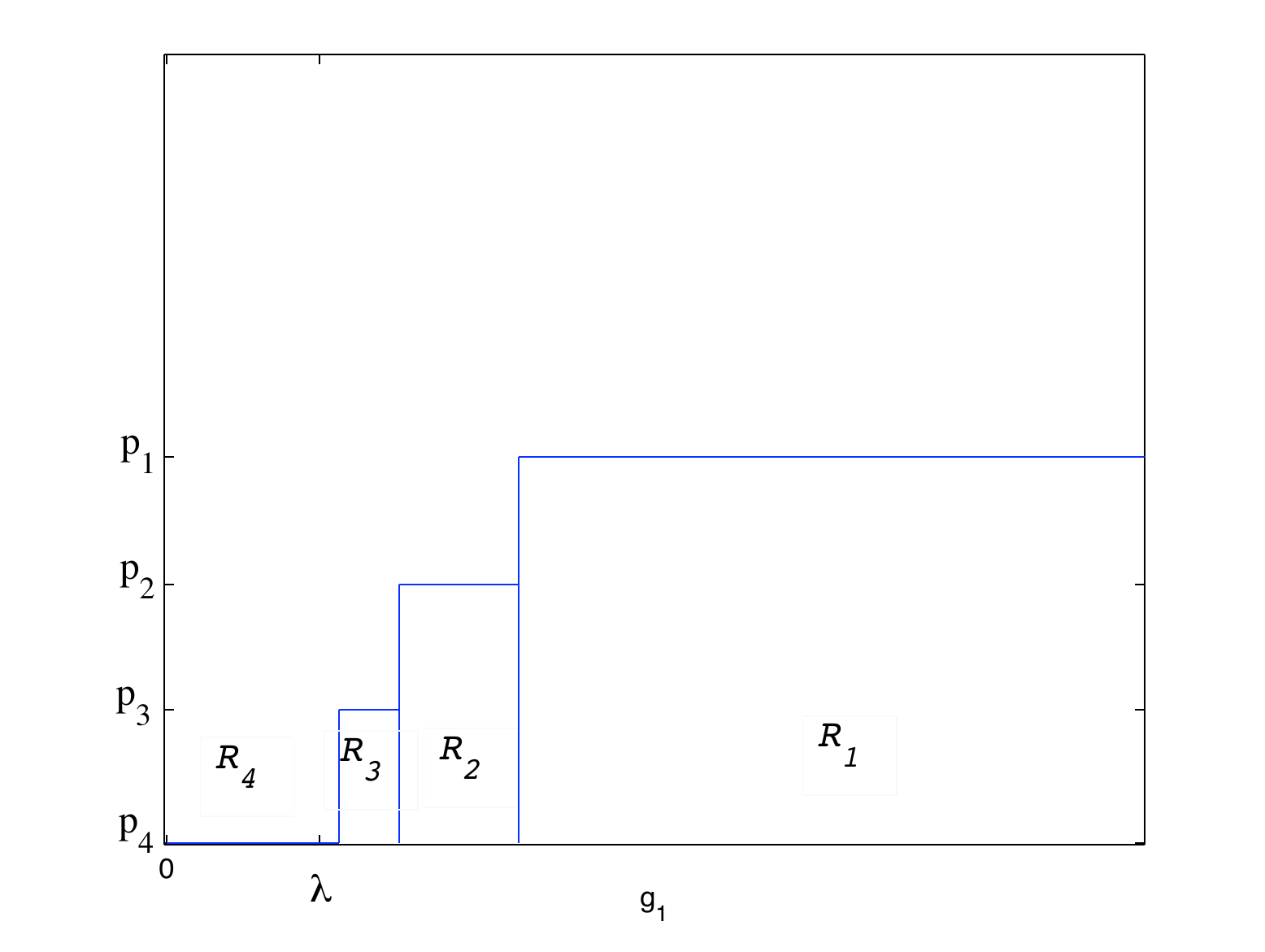}
\caption{The structure of optimum partition regions with $B=2$ bits feedback given $\lambda=1, \mu=0$}
\label{f8}
\end{figure}

\indent Next, we  present some useful properties of the optimal power solutions obtained via the modified GLA. We use the partitions ${\cal{R}}_1,\dots, {\cal{R}}_L$ and the corresponding power levels $p_1,\dots,p_L$ to denote the convergent optimal solutions.
\indent \begin{lemma}
Given partitions ${\cal{R}}_1,\dots, {\cal{R}}_L$ and the corresponding power level $p_1,\dots,p_L$, (where ${\cal{R}}_j$ and ${\cal{R}}_{j+1}, \forall j\in\{1,\dots,L-1\}$ are adjacent regions and $p_j\not=p_{j+1}$), the boundary between any two adjacent regions  ${\cal{R}}_j$ and ${\cal{R}}_{j+1} $ is given by,
$g_1=\frac{e^{(\lambda+\mu g_0)(p_j-p_{j+1})}-1}{p_j-p_{j+1} e^{(\lambda+\mu g_0)(p_j-p_{j+1})}}$
which, when $\mu\not=0$, is a monotonically increasing convex function of $g_0$  and  as $g_1\rightarrow \infty$, $g_0\rightarrow \frac{1}{\mu}(\frac{\log (\frac{p_j}{p_{j+1}})}{p_j-p_{j+1}}-\lambda)$.
\label{l2}
\end{lemma}
\indent \indent \indent \textit{Proof:}  From the NNC condition of the modified GLA,  
the boundary between two adjacent regions ${\cal{R}}_j$ and ${\cal{R}}_{j+1}$ satisfies
$\log(1+g_1p_j)-\lambda p_j-\mu g_0 p_j= \log(1+g_1p_{j+1})-\lambda p_{j+1}-\mu g_0 p_{j+1}$.
Solving the above equation for $g_1$, the result in the above Lemma follows. It is straightforward to show that it is an increasing convex function of $g_0$ by investigating 
the first and second derivatives. \\
\indent \begin{remark}
In case $\lambda > 0, \mu=0$, the AIP constraint is inactive and the ATP constraint is satisfied with equality.  In this case, the boundary between any two adjacent regions  ${\cal{R}}_j$ and ${\cal{R}}_{j+1} $ becomes
$g_1=\frac{e^{\lambda(p_j-p_{j+1})}-1}{p_j-p_{j+1} e^{\lambda(p_j-p_{j+1})}}$. 
Clearly, Problem (\ref{Q5}) reduces to an ergodic capacity maximization problem with quantized channel information. For the narrowband case, it becomes a 
 scalar quantization problem involving quantizing $g_1$ only. Note that while for the narrowband case, this no longer pertains to a cognitive radio 
 problem, the properties of the optimal quantized power allocation scheme are still important for the wideband case ($M >1$). This is due to the fact that 
 in the wideband case, it is possible that for a specific (say the $i$-th) channel, the AIP constraint is inactive ($\mu_i > 0$) while $\lambda > 0$. See 
 Section \ref{wideband} for further details.
\end{remark}
\indent \indent We now give an example to illustrate what  the optimum partition regions actually look like.   For this example, $g_0$ and $g_1$ are both exponentially distributed (Rayleigh fading) with unit mean and $L=4$ (2 bits of feedback).  The optimum partition regions are as shown in  Fig. \ref{f2} for $\lambda>0, \mu>0$, and Fig. \ref{f8} for $\lambda>0, \mu=0$. \\
\indent We obtain the following properties for the optimal quantized power levels where (as illustrated in Figure \ref{f2})
the regions ${\cal R}_1, {\cal R}_2, \ldots$ etc. are sequentially numbered, with ${\cal R}_1$ being the region closest to the 
$g_1$ axis and ${\cal R}_L$ being the region closest to the $g_0$ axis.  Note that these properties apply regardless of whether $\mu > 0$ or $\mu=0$. 
\indent \begin{theorem}
\begin{enumerate}
\item [i).] $p_1>\dots>p_L$
%\label{l3}
%\end{lemma}
%\indent \begin{lemma} 
\item [ii).] All boundaries between any two adjacent partitions  satisfy $g_1>\lambda+\mu g_0$.
%\label{l4}
%\end{lemma}
%\indent \begin{lemma}
\item [iii).] Given B bits of feedback (or $L=2^B$ regions), for the first L-1 regions, we always have strictly positive power, i.e. $p_1>\dots>p_{L-1}>0$, 
whereas  $p_L$ is simply nonnegative, i.e. $p_L \geq 0$.
%\item [iv).] As $L$ increases, the last region ${\cal{R}}_L$ becomes smaller and smaller, and as $L\rightarrow \infty$, the  boundary between $ {\cal{R}}_{L-1}$ and $ {\cal{R}}_L$
%approaches its limiting boundary $g_1=\frac{\lambda+\mu g_0}{1-(\lambda+\mu g_0) \delta^*}$, where $\delta^*=\lim_{L\rightarrow \infty} p_L$.
\item [iv).] When $\lambda+\mu \geq 1$ (note that if $\lambda=0$, $\mu \geq 1$ implies $Q_{avg}<1$, and if  $\mu=0$, $\lambda \geq 1$ corresponds to $P_{avg}<1$),  we always have $p_L = 0$. In addition, when $L$ (the number of quantized regions) is sufficiently large, no matter what $\lambda$, $\mu$ is, $p_L$ must be $0$. Additionally, as $L \rightarrow \infty$ the boundary between $ {\cal{R}}_{L-1}$ and $ {\cal{R}}_L$  approaches $g_1=\lambda+\mu g_0$ and $\lim_{L\rightarrow \infty} p_{L-1}= 0$.  
\end{enumerate}
\label{l3}
\end{theorem}
\indent \indent \indent \textit{Proof:}  See Proof in Appendices B-E.\\
\begin{remark}
The above  properties of optimal quantized power values are interesting for two reasons. From property ii), it is clear that $(g_0,g_1) \in {\cal R}_j$ for 
$j=1,2, \ldots, L-1$ satisfy the property $g_1 > \lambda + \mu g_0$ whereas for the region ${\cal R}_L$, this property may or may not be satisfied. Since the 
quantized power values in the first $L-1$ regions are strictly positive, it is easy to relate this property to the corresponding property of the full CSI based 
optimal power value which is strictly 
positive if and only if when $g_1 > \lambda^f + \mu^f g_0$.  Also, as $L \rightarrow \infty$, the boundary between ${\cal R}_{L-1}$ and ${\cal R}_L$ approaches 
$g_1 = \lambda+ \mu g_0$, thus making this relationship between the quantized power allocation scheme and the full CSI power allocation scheme stronger.

Finally, property iv) allows one to obtain an approximate quantized power allocation scheme (AQPA) for large $L$ by setting $p_L = 0$ and taking the limit as $p_{L-1} 
\rightarrow 0$. This is particularly useful as the modified GLA becomes computationally intensive for large $L$, whereas AQPA provides a performance that is extremely close to that of the modified GLA, while requiring very little computation time. A detailed description of the AQPA is provided in Section \ref{sec:aqpa} followed by 
illustrative numerical simulations in  Section \ref{sec:simul}.     
\end{remark}
\indent\indent Based on the above Lemmas,  one can solve for the optimal quantized power values given a partition ${\cal R}_1, {\cal R}_2, \ldots, {\cal R}_L$ 
 is equivalent to solving the following set of nonlinear equations for $p_1, p_2, p_3, \ldots, p_L$:
\begin{eqnarray}
E[\frac{g_1}{1+g_1p_j}-(\lambda+\mu g_0)| {\cal{R}}_j]=0,~~~~j=1,\dots,L, \; 
p_L=\max(0, p_L)
\label{nequ}
\end{eqnarray} 
where if $\mu\not=0$, $E[\frac{g_1}{1+g_1p_j}-(\lambda+\mu g_0)| {\cal{R}}_j]
=\int^{\infty}_{c_j}\int ^{r_j}_{r_{j-1}} (\frac{g_1}{1+g_1p_j}-(\lambda+\mu g_0)) f(g_0)f(g_1) dg_0 dg_1$, with $c_j=\frac{e^{\lambda(p_j-p_{j+1})}-1}{p_j-p_{j+1} e^{\lambda(p_j-p_{j+1})}}, j=1,\dots,L-1, c_{L}=0$ and $r_j=\frac{1}{\mu}(\frac{\log \frac{p_j*g_1+1}{p_{j+1}*g_1+1}}{p_j-p_{j+1}}-\lambda),  j=1,\dots,L-1, r_{0}=0, r_{L}=\infty$. When $\mu=0$, $E[\frac{g_1}{1+g_1p_j}-(\lambda+\mu g_0)| {\cal{R}}_j]=\int ^{c_{j-1}}_{c_{j}} (\frac{g_1}{1+g_1p_j}-\lambda) f(g_1) dg_1$,
with $ c_0=\infty$. (\ref{nequ}) can be solved efficiently by any suitable nonlinear equation solver. \\
\indent\indent Now that we have an algorithm based on the modified GLA for solving for the (possibly locally optimal)  quantized power values for fixed 
$\lambda, \mu$, 
 we can go back to solving the dual problem for finding the optimal values $\lambda$ and $\mu$.  To this end,  we solve the 
 associated KKT conditions (involving the average power and the average interference constraints) numerically (e.g, via a bisection method).  One can thus repeat the above two steps by solving (\ref{Q7}) and the dual problem iteratively until a satisfactory convergence criterion is met. An algorithmic format for this procedure is provided for the more general wideband ($M > 1$) case in the next subsection.
%Then we can use the following algorithm to solve the optimization problem (\ref{Q6}).\\
%\\
% \begin{tabular}{l}
%Algorithm 2\\
% \hline
%1) Let $\lambda=0$, then $\mu$ must satisfy $\mu>0$. Using bisection method to solve\\
%~~~$ \sum^L_{j=1}E[g_0 p_{j}| {\cal{R}}_j]Pr ({ \cal {R}}_j)= Q_{avg} $ for $\mu$ by running modified GLA. 
%\\~~~Then check if $\sum^L_{j=1}E[p_{j}| {\cal{R}}_j]Pr ({ \cal {R}}_j)\leq P_{avg}$ meet, corresponding \\
%~~~codebook is the solution and stop; otherwise continue.\\
%2) if not satisfy 1), then we must have $\lambda>0$.  Check If $P_{avg} \leq Q_{avg}$, $\mu$ \\
%~~~must satisfy $\mu=0$, then find $\lambda$ by solving $\sum^L_{j=1}E[p_{j}| {\cal{R}}_j]Pr ({ \cal {R}}_j)=P_{avg}$, \\
%~~~corresponding codebook $\{p_{1},\dots,p_{L}\}$ will be the solution and stop;\\~~~otherwise continue.\\
%3) if not satisfy 2), then we must have $\lambda>0, \mu>0$. Solving both 
%\\~~~$ \sum^L_{j=1}E[g_0 p_{j}| {\cal{R}}_j]Pr ({ \cal {R}}_j)= Q_{avg} $ and $\sum^L_{j=1}E[p_{j}| {\cal{R}}_j]Pr ({ \cal {R}}_j)= P_{avg}$  \\
%~~~for $\lambda$ and $\mu$, then corresponding codebook is the solution. \\
%\hline
% \end{tabular}
% \\
 \subsection{Wideband spectrum-sharing case } \label{wideband}
  \indent The above algorithm for  the narrowband case can be easily extended to the wideband case corresponding to the original problem 
  (\ref{Q1}).  For this scenario, 
the Lagrangian function becomes,
{\small
\begin{eqnarray}
&&\!\!\!\!\!\!\!\!\!\!\!\!\!\! L(P,\lambda, \textbf{u}) = \sum^L_{j=1}(\frac{1}{M}\sum^M_{i=1}E[\log(1+g^{i}_{1}p_{ij})|{\cal{R}}_j]) Pr ({ \cal {R}}_j)\nonumber\\
&&-\lambda(\sum^L_{j=1}(\frac{1}{M}\sum^M_{i=1}E[p_{ij}| {\cal{R}}_j]) Pr ({ \cal {R}}_j)- P_{avg})
-\sum^M_{i=1} \mu_i (\sum^L_{j=1}E[g^{i}_0 p_{ij}|{\cal{R}}_j] Pr ({ \cal {R}}_j)-Q^i_{avg})
\label{eq:lagwide}
\end{eqnarray}}
where $\lambda$ and $\mu_i$ are the nonnegative Lagrange multipliers
associated with the ATP constraint and $i$th AIP constraint  respectively.
The Lagrange dual function $g(\lambda, \{\mu'_i\})$ is defined as 
{
\begin{eqnarray}
\max_{p_{ij} \geq 0 ~\forall i, j} & & \frac{1}{M}{\sum^M_{i=1} }\sum^L_{j=1}E[\log(1+g^i_1p_{ij})-\lambda p_{ij}-\mu^{\prime}_i g^i_0 p_{ij}|
{\cal{R}}_j]Pr ({ \cal {R}}_j)
\label{Q11}
\end{eqnarray}} 
where $\mu^{\prime}_i=M\mu_i,~\forall i$,  and the dual problem is
$\min_{\lambda\geq 0, ~\mu^{\prime}_i \geq 0, \forall i}  g(\lambda, \{\mu'_i\})+\lambda P_{avg}+\sum^M_{i=1} \frac{\mu^{\prime}_i}{M} Q^i_{avg}$.
 
\indent Similar to the narrowband case, we first consider the problem (\ref{Q11}) to obtain $g(\lambda, \{\mu^{\prime}_i\})$ with given $\lambda$ and $\{\mu^{\prime}_i\}$.
Denote by ${\cal{R}}^i_j$  the $j$-th quantization region for the $i$-th band where $\bigcup^M_{i=1}{\cal{R}}^i_j={\cal{R}}_j$. 
Then problem (\ref{Q11}) can be decomposed into M parallel subproblems, where  for each band $i, i=1,\dots, M$
{
\begin{eqnarray}
\max_{p_{ij} \geq 0 ~\forall j}  \sum^L_{j=1}E[\log(1+g^i_1p_{ij})-\lambda p_{ij}-\mu^{\prime}_i g^i_0 p_{ij}|{\cal{R}}^i_j]Pr ({ \cal {R}}^i_j)
\label{Q13}
\end{eqnarray}} 
  is defined as the sub-dual function $g_i(\lambda, \mu^{\prime}_i)$ and $g(\lambda, \{\mu^{\prime}_i\})=\frac{1}{M}{\sum^M_{i=1} }g_i(\lambda, \mu^{\prime}_i)$. 
  This kind of duality method
is also known as the 'dual decomposition algorithm' \cite{Zhang08}.  Since each subproblem (\ref{Q13}) is similar to the problem (\ref {Q7}) for the narrowband case and can be similarly solved by using a modified GLA.   $\lambda$ and $\{ \mu^{\prime}_i\}$ can be also obtained in a manner similar to the narrowband case. These two steps are then repeated until a satisfactory convergence criterion is met. Due to the 
increased complexity resulting from the  
presence of multiple bands, we provide below a description of the overall optimization algorithm (Algorithm 1) for solving (\ref{Q5}). \\
{\bf Algorithm 1}:
\begin{enumerate}
\item  Let $\lambda=0$, then all $\mu^{\prime}_i, i=1,\dots, M$ must satisfy $\mu^{\prime}_i>0.$   Starting with some random 
initial power codebook, for each $i$, find 
$\mu^{\prime}_i$ by solving $ \sum^L_{j=1}E[g^i_0 p_{ij}| {\cal{R}}_j]Pr ({ \cal {R}}_j)= Q^i_{avg} $ and then obtain the corresponding 
(locally) optimal power codebook $\{p_{i1},\dots,p_{iL}\}$ using a modified GLA. Repeat these two steps until convergence resulting 
in a power codebook $ \{\textbf{P}_{1},\dots,\textbf{P}_{L}\} $. With this codebook, if $\sum^L_{j=1}(\frac{1}{M}\sum^M_{i=1}E[p_{ij}| {\cal{R}}_j])
 Pr ({ \cal {R}}_j)\leq P_{avg}$, it is an optimal power codebook and stop; otherwise go to step 2). 
%\item If 1) is not satisfied, then we must have $\lambda>0$.  If $P_{avg} \leq \frac{1}{M}\sum^M_{i=1} Q^i_{avg}$, 
% $\mu^{\prime}_i$ must satisfy $\mu^{\prime}_i=0,\forall i$. Then use
%M parallel modified GLA (one for each subproblem) to find the power codebook and find the corresponding $\lambda$ by solving 
%$\sum^L_{j=1}(\frac{1}{M}\sum^M_{i=1}E[p_{ij}| {\cal{R}}_j]) Pr ({ \cal {R}}_j)= P_{avg}$. Repeat these two steps until convergence 
%and the corresponding codebook
%$ \{\textbf{P}_{1},\dots,\textbf{P}_{L}\} $ will be an optimal solution and stop. Otherwise go to Step 3).
\item If 1) is not satisfied,  we must have $\lambda>0$.%and at least one $\mu^{\prime}_i>0$. 
 ~For a given  $\lambda$, for each $i$, use the modified GLA to find an optimal power codebook first with $\mu^{\prime}_i=0$. %and repeat until convergence. 
 If $ \sum^L_{j=1}E[g^i_0 p_{ij}| {\cal{R}}_j]Pr ({ \cal {R}}_j)\leq Q^i_{avg} $, then the corresponding optimal 
codebook $\{p_{i1},\dots,p_{iL}\}$ (obtained via the modified GLA) is an optimal solution for this $i$-th subproblem, otherwise, $\mu^{\prime}_i > 0$, and 
can be found by solving $ \sum^L_{j=1}E[g^i_0 p_{ij}| {\cal{R}}_j]Pr ({ \cal {R}}_j)= Q^i_{avg} $. Find the
corresponding optimal codebook entry $\{p_{i1},\dots,p_{iL}\}$ for the $i$-th subband , and then use this codebook to find an updated value of $\lambda$ by solving $\sum^L_{j=1}(\frac{1}{M}\sum^M_{i=1}E[p_{ij}| {\cal{R}}_j]) Pr ({ \cal {R}}_j)=P_{avg}$. Repeat these steps until convergence and the final codebook will 
be an optimal codebook for the wideband spectrum sharing problem (\ref{Q5}). 
\end{enumerate}

\begin{remark}
Note that it is straightforward to extend the global convergence and empirical consistency results of Theorem \ref{the2} to the wideband case. Similarly, 
Lemma  \ref{l3} also holds for the wideband case in the sense that the properties i)-iv) hold for each $\{p_{i1}, p_{i2}, \ldots p_{iL}\}, \; \forall i=1,2,\ldots, M$ 
with $\mu$ replaced by $\mu_i, i=1,2, \ldots, M$ and  $\lambda$ representing the Lagrange multiplier associated with the average sum power constraint 
in (\ref{eq:lagwide}). 
\label{re:wideprop}
\end{remark}
 \subsection{Approximate Quantized Power Allocation Algorithm (AQPA)} \label{sec:aqpa}
 \indent Although an offline algorithm, the complexity of modified GLA for determining the optimal quantized power 
  is very high for even a moderately large value of $L$. This is due to the fact that the optimal channel partitions and the 
  corresponding optimal power codebook are obtained via empirically generating a large number of channel realizations as training samples. 
  As $L$ increases, the number of training samples required will also increase. Thus motivated, we use part iv) of Lemma \ref{l3} 
   to derive a low-complexity suboptimal scheme for implementing the modified GLA for large $L$ values. Below we describe this scheme for 
   the narrowband case. A similar scheme for the wideband case can be designed accordingly.\\
\indent Note that part iv) of Lemma \ref{l3} states that as $L\rightarrow \infty$, $p_L=0$ and $p_{L-1}\rightarrow 0$. Applying these approximations to (\ref{nequ}) allows us to obtain an approximate but computationally efficient algorithm (called approximate quantized power allocation algorithm (AQPA)) for large $L$. AQPA first solves $E[\frac{g_1}{1+g_1p_{L-1}}-(\lambda+\mu g_0)| {\cal{R}}_{L-1}]=0$ for $p_{L-2}$ by substituting $p_L=0$ and taking the limit $p_{L-1}\rightarrow 0$, which,  
if $\mu > 0$,  is equivalent  to solving $\int^{\infty}_{\lambda}\int ^{\frac{g_1-\lambda}{\mu}}_{\frac{1}{\mu}(\frac{\log(1+g_1 p_{L-2})}{p_{L-2}}-\lambda)} (g_1-(\lambda+\mu g_0)) f(g_0)f(g_1) dg_0 dg_1=0$ for $p_{L-2}$. When $\mu=0$, it is equivalent to  solving for  $p_{L-2}$ from$ \int ^{\frac{e^{\lambda p_{L-2}}-1}{p_{L-2}}}_{\lambda} (g_1-\lambda) f(g_1) dg_1=0$.  Note that the above equations (for both $\mu > 0$ and $\mu=0$) involve only one 
variable: $p_{L-2}$ and are thus straightforward to solve. One can then recursively compute $p_{L-3}, p_{L-4}, \ldots,$ by using 
the optimality conditions for the regions ${\cal R}_{L-2}, {\cal R}_{L-3}, \ldots, $ respectively, in the reverse
direction. These equations can be solved by appropriate nonlinear equation solvers and do not require the use of large 
number of training samples. Thus AQPA is  significantly faster than GLA and  is applicable 
to the case of large number of feedback bits. 
Note however, as this is an approximate algorithm only, the performance of this algorithm becomes comparable to modified GLA only for 
large values of $L$. Numerical results presented in the next section illustrate that AQPA performs extremely well for $L \geq 16$. 

\section{Optimum Quantized Power Allocation with Noisy Limited Feedback} \label{noisyfeedback}
\indent In the previous section, we assumed ideal error-free feedback in the limited feedback model.  However, feedback channel noise can result in unavoidable erroneous feedback, which can cause the SU-TX incorrectly selecting an incorrect transmission strategy and thus dramatically degrade the capacity performance. In this section, we allow noise in the limit feedback channel model and study the ergodic capacity maximization problem (\ref{Q5}) with noisy limited feedback. 
The noisy feedback link, assumed to be memoryless,  is characterized by the index transition probabilities $\rho_{kj}, (k,j=1,\dots,L)$, which is defined as the probability of  receiving index $k$ at the SU-TX, given index $j$ was sent from the CR service provider/band manager.  Given $B=\log_2 L$ bits feedback,  denote binary representation of index $k$ and $j$ as $k_1k_2\dots k_B$ and $j_1j_2\dots j_B$ respectively, where $k_n,j_n\in\{0,1\} $ for  $n=1,\dots B $, and $k_1, j_1$ represent the 
most significant bit. We model the noisy feedback channel as $B$ independent uses of a binary symmetric channel with crossover probability  $q_f$  for every feedback bit.
Since bit errors are used to be independent, $\rho_{kj}=\prod^B_{n=1} \rho_{k_nj_n}=q_f^{d_{k,j}} (1-q_f)^{B-d_{k,j}},$ where $d_{k,j}$ is the Hamming distance between the binary representations of $k$ and $j$  \cite{Ekbatani09}\cite{Zeger90}.  \\
\indent Thus problem (\ref{Q5}) with noisy limited feedback can be reformulated as
\begin{eqnarray}
&&\max_{p_{ik} \geq 0, \forall  i, k, { \cal {R}}_j, \forall j }~  \sum^L_{j=1}\sum^L_{k=1}(\frac{1}{M}\sum^M_{i=1}E[\log(1+g^{i}_{1}p_{ik})| {\cal{R}}_j]) \rho_{kj} 
Pr ({ \cal {R}}_j)\nonumber\\
&&\sum^L_{j=1}\sum^L_{k=1}E[g^{i}_0 p_{ik}| {\cal{R}}_j] \rho_{kj} Pr ({ \cal {R}}_j)\leq Q^i_{avg},    ~\forall  i, \;
\sum^L_{j=1}\sum^L_{k=1}(\frac{1}{M}\sum^M_{i=1}E[p_{ik}| {\cal{R}}_j]) \rho_{kj} Pr ({ \cal {R}}_j)\leq P_{avg}
\label{QQ5}
\end{eqnarray} 
Note that the binary codewords representing the feedback indices for a power codebook of size $L$ can  be designed in $L!$ different ways. Thus finding the optimal  index assignment can be done by an exhaustive search for small $B$. For large $B$, one could resort to some low-complexity suboptimal index assignment schemes like \cite{Zeger90}. Note that such index reassignment schemes will yield the same codebook but with its power vectors in different location \cite{Zeger90}.  Here, given a fixed index assignment scheme, we simply concentrate on finding the optimum CSI partitions ${ \cal {R}}_j, \forall j $ and power codebook $\cal {P}$ that jointly optimizes the ergodic capacity of SU under the long term  average transmit power constraint and average interference constraint given by (\ref{QQ5}). \\
\indent Again, to keep things simple, we look at narrowband spectrum-sharing case (M=1). Using the simplified notations  $p_{j},\; j=1,2,\ldots, L$, and $ g_1, g_0, Q_{avg}$, we
  write the Lagrangian for the problem (\ref{QQ5}) with $M=1$ as 
{\begin{eqnarray}
 L(P,\lambda, \mu) =\sum^L_{j=1} \sum^L_{k=1} E[\log(1+g_1p_k)-\lambda p_k-\mu g_0 p_k |{\cal{R}}_j]\rho_{kj} Pr ({ \cal {R}}_j)+\lambda P_{avg}+\mu Q_{avg}
\end{eqnarray}}
where $\lambda$ and $\mu$ are the nonnegative Lagrange multipliers
associated with the ATP constraint and AIP constraint respectively.
Thus the Lagrange dual function $g(\lambda, \mu)$ is defined as 
\begin{eqnarray}
\max_{p_k \geq 0,~\forall k, {\cal {R}}_j,~\forall j}  \sum^L_{j=1}\sum^L_{k=1} E[\log(1+g_1p_k)-\lambda p_k-\mu g_0 p_k |{\cal{R}}_j] \rho_{kj} Pr ({ \cal {R}}_j)
\label{QQ7}
\end{eqnarray} 
and the corresponding dual problem is $\min_{\lambda\geq 0, ~\mu\geq 0}  g(\lambda, \mu)+\lambda P_{avg}+\mu Q_{avg}$. \\
\indent  We can solve the optimization problem (\ref {QQ7}) with fixed $\lambda$ and $\mu$ using another modified GLA, 
(termed as modified GLA-2 to distinguish it from the noise free case) by repeating the following two steps until convergence: 1) Using large number of training samples for $(g_0,g_1)$, assign individual $(g_0,g_1)$ samples to  
 ${\cal R}_j$ if $ \sum^L_{k=1} (\log(1+g_1p_k)-\lambda p_k-\mu g_0 p_k ) \rho_{kj}>\sum^L_{k=1} (\log(1+g_1p_k)-\lambda p_k-\mu g_0 p_k) \rho_{kn}, n=1,\dots L, n\neq j$, $\forall j=1,\dots, L$. 2) Given a partition, the optimal power codebook is given by solving the convex optimization problem 
 $\text{argmax}_{p_k\geq 0} \sum^L_{j=1} E[\log(1+g_1p_k)-\lambda p_k-\mu g_0 p_k|{\cal{R}}_j]\rho_{kj} Pr ({ \cal {R}}_j)$, 
 $\forall k=1,2,\ldots, L$.  One can then obtain the optimal power as $\max(p_k^*, 0)$, where $p_k^*$ is the solution to the equation $\sum^L_{j=1}E[\frac{g_1}{1+g_1p_k}-(\lambda+\mu g_0)| {\cal{R}}_j]\rho_{kj}=0$. \\
\indent For this power codebook, the optimal values $\lambda$ and $\mu$  can then be obtained numerically 
by solving the associated KKT conditions. One can repeat the modified GLA-2 and the algorithm for finding $\lambda, \mu$  iteratively until a satisfactory convergence criterion is met. The extension to the wideband case is obvious and is thus omitted.

\section{Numerical Results} \label{sec:simul}
\begin{figure}[h]
\centering
\includegraphics[scale=0.9]{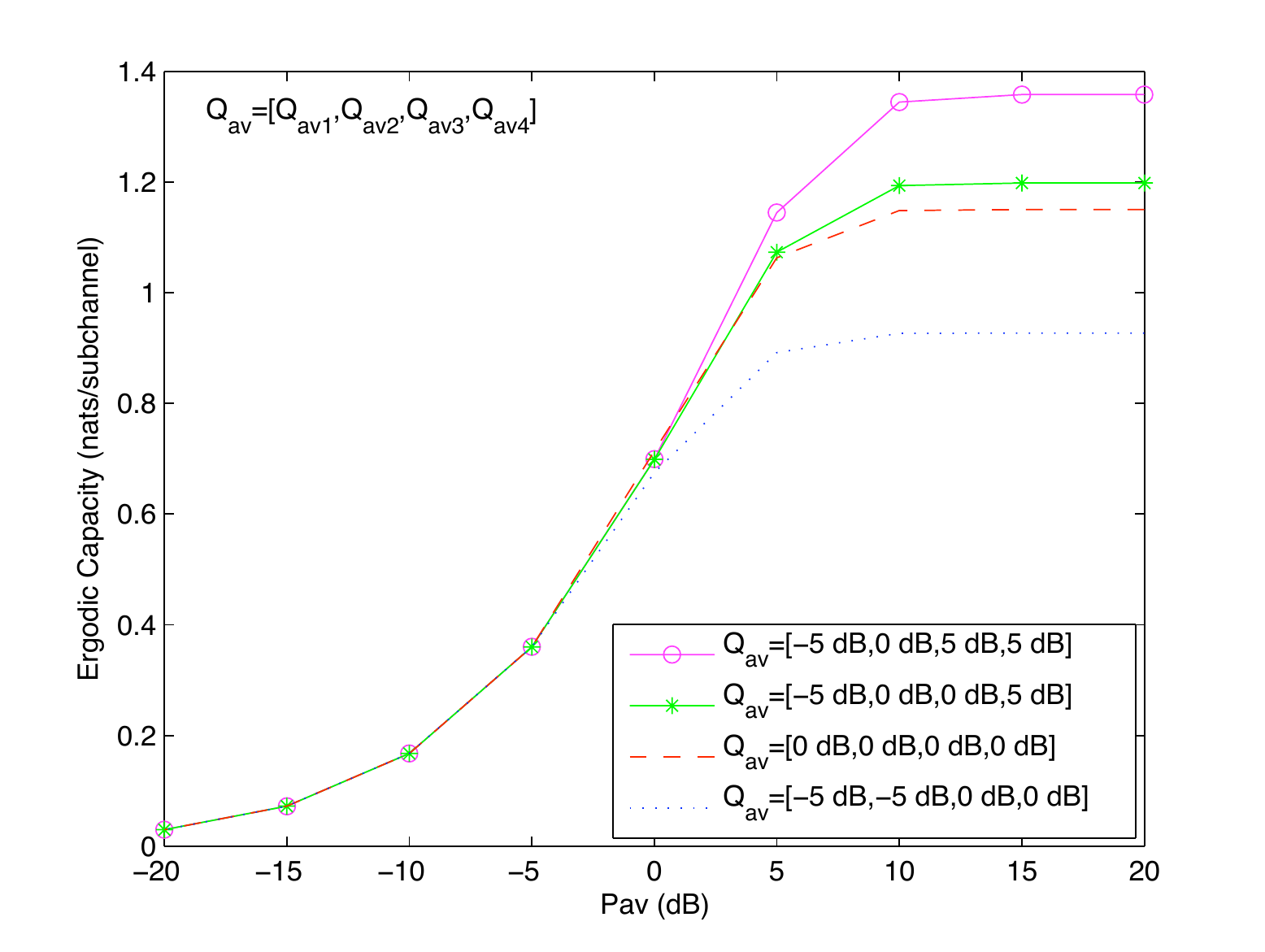}
\caption{SU ergodic capacity for spectrum sharing with 4 PUs and prefect CSI at SU-TX}
\label{f3}
\end{figure}

\begin{figure}[h]
\centering
\includegraphics[scale=0.9]{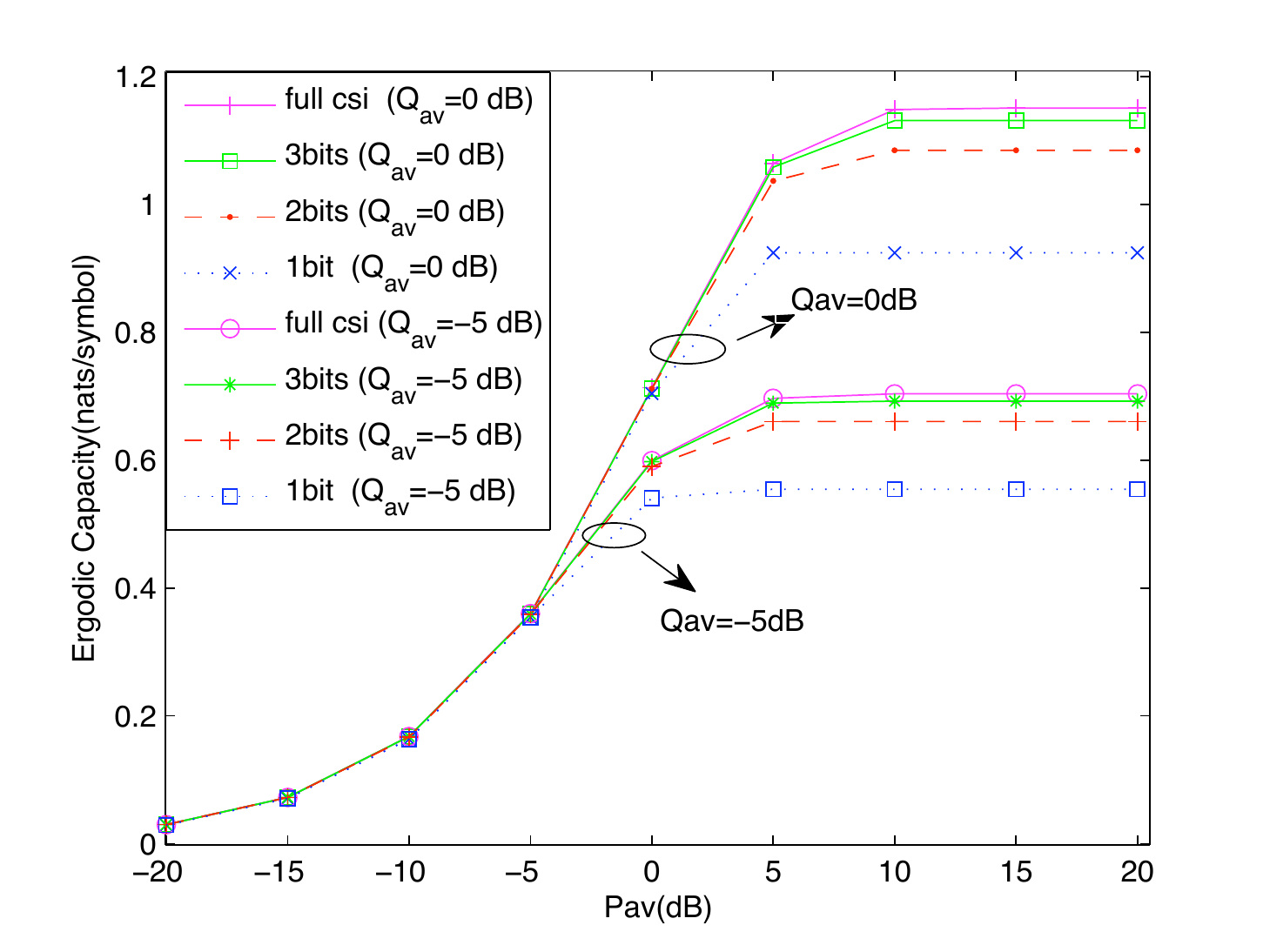}
\caption{SU Ergodic capacity with quantized power allocation (using GLA) with one PU for $Q_{av} = -5$ dB and $Q_{av} = 0$ dB }
\label{f4}
\end{figure}

\begin{figure}[h]
\centering
\includegraphics[scale=0.96]{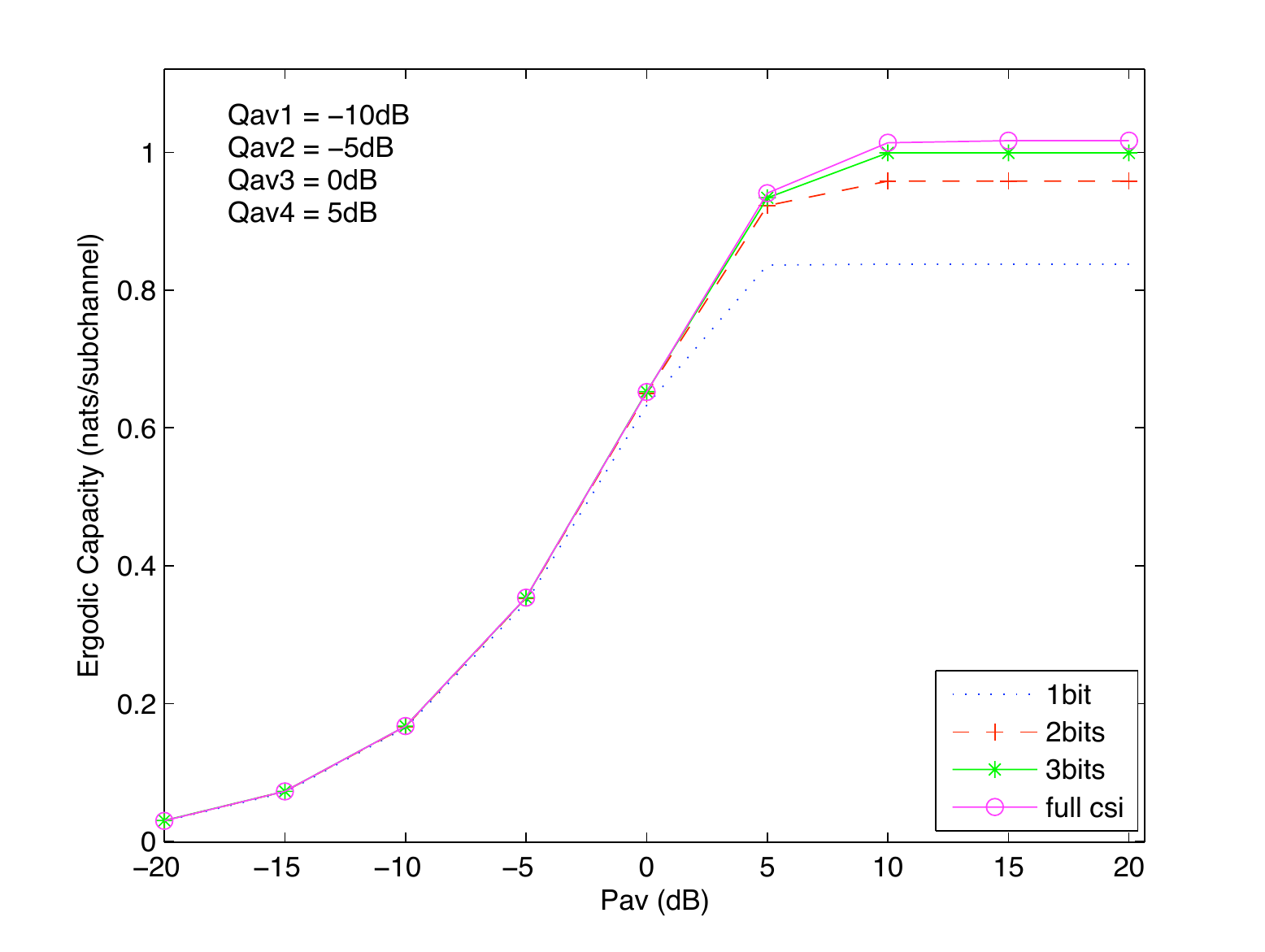}
\caption{SU Ergodic capacity with quantized power allocation (GLA) with four PUs ($M=4$)}
\label{f5}
\end{figure}

\begin{figure}[h]
\centering
\includegraphics[scale=0.96]{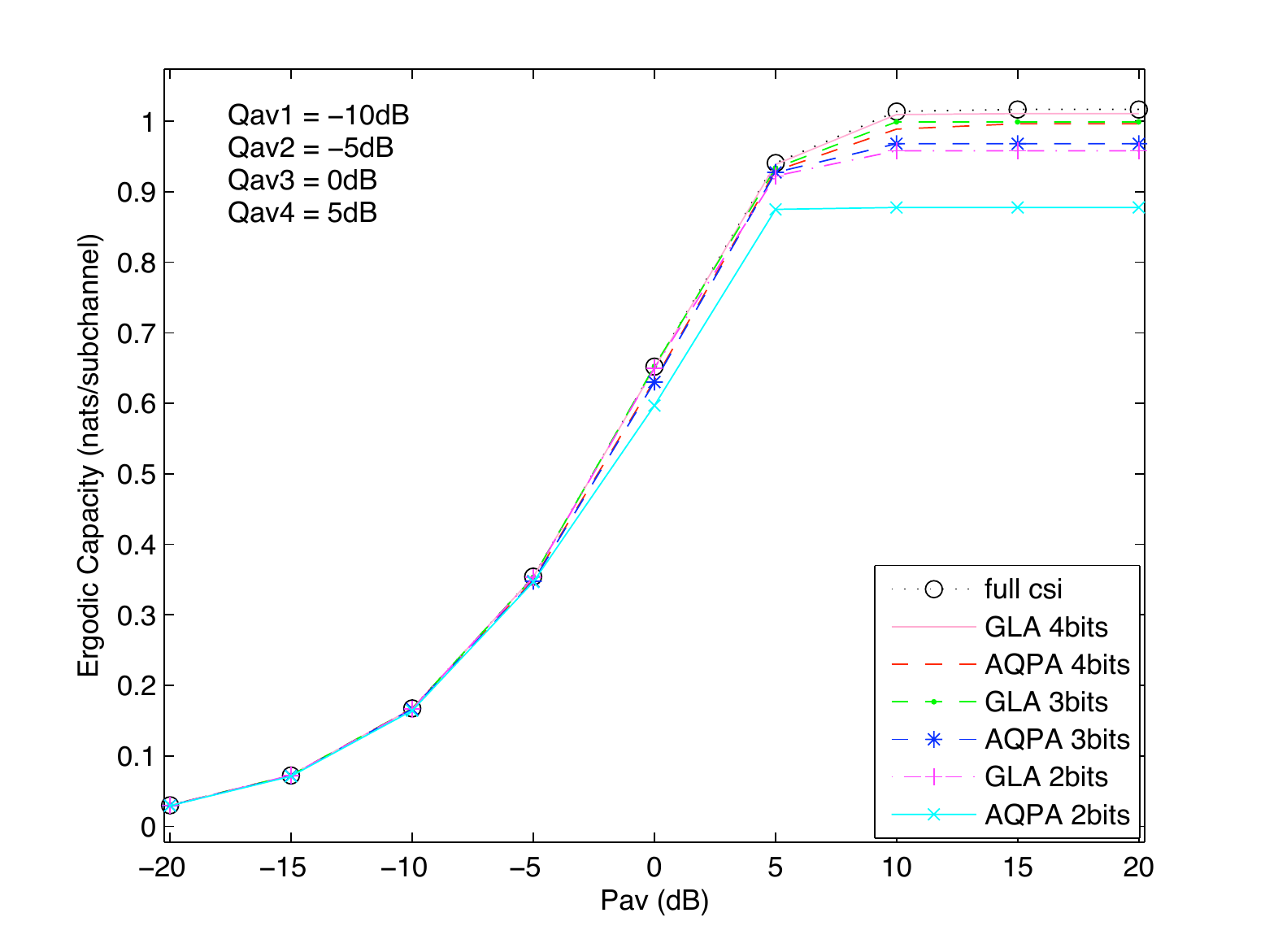}
\caption{Capacity performance of AQPA with four PUs ($M=4$)}
\label{f6}
\end{figure}

\begin{figure}[h]
\centering
\includegraphics[scale=0.96]{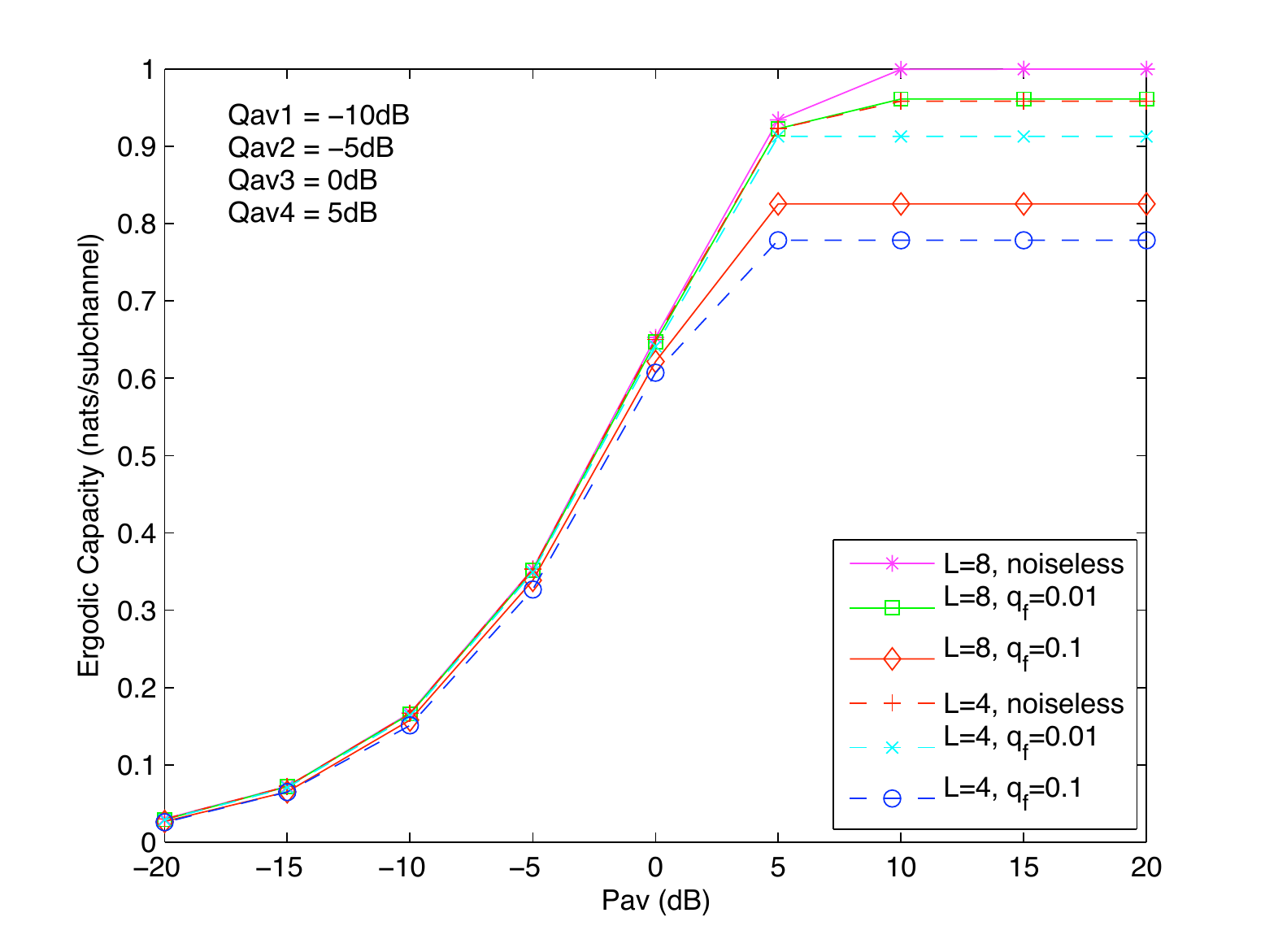}
\caption{Capacity performance of noisy limited feedback with four PUs (M=4) and different BSC crossover probabilities}
\label{f9}
\end{figure}

\indent In this section, we will evaluate the performance of the designed power allocation strategies via numerical simulations.
We implement a wideband spectrum sharing system with one SU and $M$ independent frequency bands (each band is originally licensed to a PU), where all the channels involved are assumed 
to undergo Rayleigh fading, namely  all $\textbf {g}_0$ and $\textbf {g}_1$ are exponentially distributed with unit mean. For each 
simulation, 100,000 randomly generated channel realizations for each $\textbf {g}_0$ or $\textbf {g}_1$ are used.\\
\indent Fig. \ref{f3} shows  with prefect CSI, the capacity performance of SU-TX, which shares spectrum with four PUs (M=4), with four different AIP constraints thresholds, i.e, ($Q_{av1}, Q_{av2}, Q_{av3}, Q_{av4}$)=($-5$ dB, $-5$ dB, $0$ dB, $0$ dB), ($Q_{av1}, Q_{av2}, Q_{av3}, Q_{av4}$)=($0$ dB, $0$ dB, $0$ dB, $0$ dB), ($Q_{av1}, Q_{av2}, Q_{av3}, Q_{av4}$)=($-5$ dB, $0$ dB, $0$ dB, $5$ dB) and ($Q_{av1}, Q_{av2}, Q_{av3}, Q_{av4}$)=($-5$ dB, $0$ dB, $5$ dB, $5$ dB). An interesting observation from Fig. \ref{f3} is that when $P_{av}$ is small ($P_{av}\leq -5$ dB), no matter what the value of ($Q_{av1}, Q_{av2}, Q_{av3}, Q_{av4}$) is,  the capacity performance of four curves are almost indistinguishable. This is due to the fact that (see Theorem {\ref{t1}}),  when $P_{av}\leq \min(Q_{av1}, Q_{av2},Q_{av3}, Q_{av4}$) (since $\textbf {g}_1$ is i.i.d),  all AIP constraints become inactive.  As the value of $P_{av}$ increases, the capacity performance with different ($Q_{av1}, Q_{av2}, Q_{av3}, Q_{av4}$) gradually becomes distinguishable, since in this case, the ATP and at least one AIP constraint are effective. %As expected, the capacity performance with large ($Q_{av1}, Q_{av2}, Q_{av3}, Q_{av4}$) outperform the ones with low AIP thresholds. 
However, as $P_{av}$ increases beyond a certain threshold, the capacity curves start to 
saturate, due to the fact that when $P_{av}\geq \frac{1}{4}\sum^4_{i=1} E[(\frac{1}{\mu_i g^i_0}-\frac{1}{g^i_1})^+]$,  where $\mu_i$ is given by solving $E[g^i_0 (\frac{1}{\mu_i g^i_0}-\frac{1}{g^i_1})^+] = Q^i_{av} $, only the AIP constraints are active. Thus no matter how $P_{av}$ changes, if ($Q_{av1}, Q_{av2}, Q_{av3}, Q_{av4}$) are fixed, the capacity will be unchanged. A similar observation for a narrowband spectrum sharing model with full CSI was made in \cite{Kang09}. One should note that theoretically,
the ATP corresponding to the optimal power allocation law maximizing the SU ergodic capacity over a Rayleigh fading channel under an AIP constraint with perfect CSI is infinity \cite{Musavian092}.  Since here we use large numbers of randomly generated channel realizations samples in the  simulation studies,  the ATP for maximizing SU ergodic capacity under an AIP constraint is large but not infinite.   \\
%\begin{figure}[h]
%\centering
%\includegraphics[scale=0.5]{f3}
%\caption{Capacity performance for SU-TX in only one PU case with prefect CSI  obtained by Algorithm 1.}
%\label{f3}
%\end{figure}
\indent Fig. \ref{f4} shows the capacity performance of SU sharing a narrowband spectrum with one PU with limited feedback for $Q_{av}=  -5$ dB and $Q_{av}=  0$ dB respectively, and illustrates the effect of increasing the number of feedback bits on the capacity performance. For comparison, we also plot the corresponding capacity performance with full CSI. The striking observation from Fig. \ref{f4} is that introducing one extra bit of feedback substantially reduces the gap with capacity based on perfect CSI.  This property is not very obvious when $P_{av}$ is small, for example when $P_{av}\leq -5$ dB ($P_{av}\leq 0$ dB) for $Q_{av}=  -5$ dB ($Q_{av}=  0$ dB). But with increasing $P_{av}$, it becomes more pronounced. To be specific,  for  $Q_{av}=  -5$ dB case, at $P_{av}=10$ dB, with $1$ bit, $2$ bits and $3$ bits of feedback, the percentage capacity loss is approximately  $21.23 \%, 6.21\%$ and $1.62 \%$ respectively, and for both $Q_{av}=  -5$ dB and $Q_{av}=  0$ dB cases, only 3 bits feedback can result in secondary ergodic capacity very close to that with  full CSI. This is very encouraging since only a 
small number of bits of feedback are required to achieve close 
performance to the full CSI case. It can be also seen  that the capacity performance with large AIP threshold ($Q_{av}=  0$ dB) outperform the ones with low AIP threshold ($Q_{av}=  -5$ dB), as expected.  A similar behaviour can be also observed in Fig. \ref{f5} for a wideband spectrum sharing case ($M=4$)(($Q_{av1}, Q_{av2}, Q_{av3}, Q_{av4}$)=($-10$ dB, $-5$ dB, $0$ dB, $5$ dB)). \\
\indent In Fig. \ref{f6} we compare the performance of AQPA with modified GLA, where SU shares the spectrum with four PUs $(M=4)$ and AIP constraint thresholds ($Q_{av1}, Q_{av2}, Q_{av3}, Q_{av4}$)=($-10$ dB, $-5$ dB, $0$ dB, $5$ dB).  It illustrated that with the same number of bits of feedback, the gap between AQPA and modified GLA becomes smaller as $L$ increases. For example, when $P_{av}=15$ dB, the capacity loss by using AQPA instead of GLA is about $8.38 \%, 3.12\%$ and $1.42 \%$ for 2 bits, 3 bits and 4 bits feedback respectively. It is clearly seen that AQPA with 4 bits feedback can almost approach the full CSI performance. It is also noticed that 
for a fixed $\lambda$ and $\mu$ with $M=4$ and 4 bits of feedback, AQPA is approximately 10 times faster than GLA operating with 100,000 training samples on a 
Pentium 3 processor. \\
\indent Finally, we investigate SU ergodic capacity performance with noisy limited feedback in Fig. \ref{f9}, for a wideband spectrum sharing case ($M=4$ and ($Q_{av1}, Q_{av2}, Q_{av3}, Q_{av4}$)=($-10$ dB, $-5$ dB, $0$ dB, $5$ dB)). It can be observed that as the feedback becomes less reliable (the crossover probability $q_f$ increases), significant capacity performance degradation occurs, especially in high $P_{avg}$. For example, when $P_{avg}=10$ dB, for 3 (2) bits feedback, a noisy feedback channel with $q_f=0.01$ and $q_f=0.1$ can result in approximately $3.843 \%$ ($4.769\%$) and $17.394 \%$ ($18.783 \%$) capacity loss respectively, compared to the noise-free case. This clearly illustrates that as the quality of feedback link degrades, the benefit of designing an optimal power codebook diminishes rapidly.

\section{Conclusions and extensions} \label{conclu}
We have derived quantized power allocation algorithms for a wideband spectrum sharing system with one secondary user and multiple primary users, each licensed to 
use  a separate frequency band, each band modelled as independent block fading channels. 
The objective has been to maximize the SU ergodic capacity under an average sum transmit power constraint and individual average interference constraints at the PU receivers. Modified Generalized Lloyd-type algorithms (GLA) have been derived and various properties of the 
quantized power allocation laws have been presented, along with a rigorous convergence and consistency proof of the modified GLA based algorithm. 
By appropriately exploiting the properties of the quantized power values for large number of bits of feedback, we have also derived approximate quantized power allocation 
algorithms that perform very close to the modified GLA based algorithms but are significantly faster. Finally, we have presented an extension of the modified GLA based quantized power allocation algorithm to the case of noisy feedback channels. 
   Future work will include
deriving expressions for asymptotic (as the number of feedback bits goes to infinity) capacity loss with quantized power allocation, 
consideration of primary interference at the secondary receiver and designing of optimal index assignment schemes for quantized power allocation 
with noisy limited feedback. 
\begin{appendix}
\subsection{Proof of Theorem \ref{t1}}
\indent \textbf{1)} Note that the Karush-Kuhn-Tucker (KKT) conditions are necessary and sufficient for a convex optimization problem. This implies that all 
the conditions stated in Theorem \ref{t1} are necessary and sufficient. When $\lambda^f=0$,  from the complementary slackness condition, the constraint $\frac{1}{M}\sum^M_{i=1}E[p_i(\textbf {g}_0, \textbf {g}_1)] \leq P_{avg}$ does not come into play.  In this case,  
the optimization problem (\ref{Q1}) becomes M completely independent parallel subproblems all having the same structure:
\begin{eqnarray}
&&\max_{p_i(\textbf {g}_0, \textbf {g}_1) \geq 0} ~E[\log(1+g^i_1p_i(\textbf {g}_0, \textbf {g}_1))]\nonumber\\
&&~~~~s. t. ~~~~~E[g^i_0 p_i(\textbf {g}_0, \textbf {g}_1)] \leq Q^i_{avg}, \; \forall i=1,2,\ldots, M
\label{Q23}
\end{eqnarray} 
and it is easy to verify that in the above optimization problem, each constraint holds with equality, namely $E[g^i_0 p_i(\textbf {g}_0, \textbf {g}_1)] = Q^i_{avg}~~\forall i$. Thus for each $i$, from the complementary slackness condition, one can easily show that  
 $\mu^f_i > 0$.  Hence, in this case, we have the optimal solution
 \begin{eqnarray}
p_i^*(\textbf {g}_0, \textbf {g}_1)=(\frac{1}{\mu^f_i g^i_0}-\frac{1}{g^i_1})^+  ~~~~~~ \forall i
\end{eqnarray} 
where $\mu^f_i$ is determined such that $E[g^i_0 (\frac{1}{\mu^f_i g^i_0}-\frac{1}{g^i_1})^+] = Q^i_{avg}  ~~\forall i$. From feasibility, we also have $\frac{1}{M}\sum^M_{i=1}E[(\frac{1}{\mu^f_i g^i_0}-\frac{1}{g^i_1})^+] \leq P_{avg}$.\\
\indent \textbf{2)} When $\lambda^f>0$,  again from the complementary slackness condition,
$\frac{1}{M}\sum^M_{i=1}E[p_i(\textbf {g}_0, \textbf {g}_1)] = P_{avg}$. 
\indent \indent $\bullet$  If $ \mu^f_i> 0$, then corresponding AIP constraint must satisfy with equality ($E[g^i_0 p_i(\textbf {g}_0, \textbf {g}_1)] = Q^i_{avg}$)
and hence  the optimal solution for the $i$-th channel is 
\begin{eqnarray}
p_i^*(\textbf {g}_0, \textbf {g}_1) =(\frac{1}{\lambda^f+\mu^f_i g^i_0}-\frac{1}{g^i_1})^+
\end{eqnarray} 
where $\mu^f_i$ is determined from $E[g^i_0 (\frac{1}{\lambda^f+\mu^f_i g^i_0}-\frac{1}{g^i_1})^+] = Q^i_{avg}$ given $\lambda^f$.\\
\indent \indent $\bullet$  If $ \mu^f_i= 0$, then the corresponding AIP constraint satisfies $E[g^i_0 p_i(\textbf {g}_0, \textbf {g}_1)] \leq Q^i_{avg}$, and
in this case  the optimal solution for the $i$-th channel is 
\begin{eqnarray}
p_i^*(\textbf {g}_0, \textbf {g}_1) =(\frac{1}{\lambda^f}-\frac{1}{g^i_1})^+
\end{eqnarray} 
In this scenario we also have $E[g^i_0 (\frac{1}{\lambda^f}-\frac{1}{g^i_1})^+]=E[(\frac{1}{\lambda^f}-\frac{1}{g^i_1})^+] \leq Q^i_{avg}$, since $\textbf {g}_0 $ and $ \textbf {g}_1 $ are independent,  and $E[g_0^i] = 1,\forall i$.\\
\indent Thus when $\lambda^f>0$, the optimal solution is given by
\begin{eqnarray}
p_i^*(\textbf {g}_0, \textbf {g}_1) =\begin{cases}
(\frac{1}{\lambda^f}-\frac{1}{g^i_1})^+   ~~~~~~~\text{if}~~~  E[(\frac{1}{\lambda^f}-\frac{1}{g^i_1})^+] \leq Q^i_{avg}\\
(\frac{1}{\lambda^f+\mu^f_i g^i_0}-\frac{1}{g^i_1})^+~~~~~~~~~~~otherwise\\
\end{cases}
\label{Q3}
\end{eqnarray}
where $\lambda^f$ is determined such that $\frac{1}{M}\sum^M_{i=1}E[p_i] = P_{avg}$.\\

\subsection{Proof of Theorem \ref{the2}} \label{appen1}
\indent \indent \indent \textit{Proof:}  For the  modified GLA, one can define a distortion measure  $d((g_0, g_1), p)=-(\log(1+g_1p)-\lambda p-\mu g_0 p)$.  For such non-difference distortion measures,  following \cite{Linder99},  one can ensure nonnegativity of the distortion measure by introducing a  modified distortion measure as $\hat {d}((g_0, g_1), p)=d((g_0, g_1), p)-\min_p  d((g_0, g_1), p)$.  Since $d((g_0, g_1), p)$ is a convex function of $p$ for fixed $(g_0,g_1)$, we get the unique minimum $p*=(\frac{1}{\lambda+\mu g_0}-\frac{1}{g_1})^+$, thus $\min_p  d((g_0, g_1), p)= d((g_0, g_1), p*)$.  Therefore we have $\hat {d}((g_0, g_1), p)\geq0$.  Since $ d((g_0, g_1), p*)$ is constant for a given $(g_0, g_1)$, thus using distortion measure  $\hat {d}((g_0, g_1), p)$ instead of $d((g_0, g_1), p)$ does not affect the results of modified GLA. One can easily show that $\hat{d}$ satisfies the following properties:
(1) $\hat{d}$ is continuous and $\hat{d}\in[0, \infty)$,
(2)$\hat{d}((g_0, g_1), p)$ is a convex function of $p$ for each fixed $(g_0, g_1)$, 
(3) for each $(g_0, g_1), \hat{d}(\widetilde{(g_0, g_1)}, p) \rightarrow \infty$, as $\widetilde{(g_0, g_1)} \rightarrow  (g_0, g_1) $ and $\parallel p\parallel  \rightarrow \infty$, and (4) the partition boundaries in the channel space $(g_0,g_1)$ have zero probability.

%$\int\int_{(g_0, g_1): d((g_0, g_1), p_i)=d((g_0, g_1), p_j)} dg_0 dg_1=0$ for $i\not=j \in \{1,\dots,L\}$ and $p_i\not =p_j$.\\
%It is obvious that function $\hat{d}$ satisfies 1); Since $\frac{\partial^2{\hat{d}}}{\partial p^2}=\frac{(g_1)^2}{(1+g_1p)^2}\geq0$, thus $\hat{d}$ satisfies  2);  $\hat{d}$ meets the property  3), due to 
%\begin{eqnarray}
%&&\lim_{\parallel p\parallel \rightarrow \infty} -(\log(1+g_1p)-\lambda p-\mu g_0 p)- d((g_0, g_1), p*)\nonumber\\
%&&=\lim_{\parallel p\parallel \rightarrow \infty} p(\frac{-\log(1+g_1p)}{p}+\lambda+\mu g_0-\frac{ d((g_0, g_1), p*)}{p}) \nonumber\\
%&&=(\lim_{\parallel p\parallel \rightarrow \infty} p)(\lim_{\parallel p\parallel \rightarrow \infty}(\frac{-\log(1+g_1p)}{p})+\lambda+\mu g_0-0) \nonumber\\
%&&=(\lim_{\parallel p\parallel \rightarrow \infty} p)(\lim_{\parallel p\parallel \rightarrow \infty}(\frac{-g_1}{1+g_1p})+\lambda+\mu g_0) \nonumber\\
%&&=(\lim_{\parallel p\parallel \rightarrow \infty} p)(\lambda+\mu g_0) \nonumber\\
%&&=\infty
%\end{eqnarray}
%where $\lambda+\mu g_0\not=0$, since $\lambda$ and $\mu $ can not both be zero at the same time. 
Properties 1), 2) and 3) are easy to show and the proofs here are omitted.
Property 4) holds due to the assumption of 
continuous fading channels in this work. Note that this is also a necessary condition for a codebook to be optimal for a given partition \cite{Gersho92}.  Note also that 
the popular fading distributions such as Rayleigh, Rician and Nakagami and Log-normal etc. all satisfy  the absolutely continuity assumption. It is then easy to show that for these types of fading scenarios, the cumulative distribution function (cdf) of $(g_0,g_1)$, denoted by $F$,  satisfies the following properties \cite{Sabin86}:
(5)  F contains no singular-continuous part and (6)  $\int  \hat{d}((g_0, g_1), p) dF(g_0, g_1)<\infty $ for each $p$  (implying a finite average distortion).
\indent Next, let $\textbf {g}$ denote $(g_0, g_1)$. Noting that $\{\textbf {g}(\omega)\}$ is a stationary ergodic sequence with a cdf $F$, and letting $F_{n,\omega}$ be the empirical distribution function of the first n members of the sequence \cite{Sabin86},  one can show that 
 for almost every $\omega$, $\{F_{n,\omega}\}$ and $F$ satisfy (see Lemma 4 of \cite{Sabin86}) (7) $\{F_{n}\}$ converges weakly to the  $F$ and (8)  
 $ \lim_n \int  \hat{d}((g_0, g_1), p) dF_n(g_0, g_1)=  \int  \hat{d}((g_0, g_1), p) dF(g_0, g_1), \; {\text {for every}} \: p $.
 
Hence, from  \cite{Sabin86}, we can conclude that the modified GLA satisfies properties 1) to 8). Therefore, Lemmas 1-3 of \cite{Sabin86}
are applicable to the modified GLA with probability one and the modified GLA satisfies the global convergence and empirical consistency properties as defined in 
\cite{Sabin86}. 

\subsection{Proof of Theorem \ref{l3} i)} \label{appen2}
\indent \indent \indent \textit{Proof:}  We need to prove that for any two adjacent  regions ${\cal{R}}_j$ and ${\cal{R}}_{j+1}, j=1,\dots,L-1$,  $p_j>p_{j+1}$. 
 Given an arbitrary $g_0$  satisfying  $0\leq g_0<\frac{1}{\mu}(\frac{\log (\frac{p_j}{p_{j+1}})}{p_j-p_{j+1}}-\lambda)$ (assuming $\mu > 0$), suppose  there is a point $(g_0, g^a_{1})\in {\cal{R}}_j$ and a point $(g_0, g^c_{1})\in {\cal{R}}_{j+1}$ (neither of these two points is on the boundary),  and let $(g_0, g^b_{1})$ denote the point on the boundary corresponding to the same $g_0$, which from Lemma \ref{l2}, is given by
$g^b_1=\frac{e^{(\lambda+\mu g_0)(p_j-p_{j+1})}-1}{p_j-p_{j+1} e^{(\lambda+\mu g_0)(p_j-p_{j+1})}}$
Then, we have  $g^a_1> g^b_1> g^c_1$. Now suppose $p_j<p_{j+1}$.   Since  $(g_0, g^a_{1})\in {\cal{R}}_j$, we have
$\log(1+g^a_1p_j)-\lambda p_j-\mu g_0 p_j\geq \log(1+g^a_1p_{j+1})-\lambda p_{j+1}-\mu g_0 p_{j+1}$
 As $p_j<p_{j+1}$, we have
 \begin{eqnarray}
 (\lambda +\mu g_0)(p_{j+1}-p_j)&\geq& \log(\frac{1+g^a_1p_{j+1}}{1+g^a_1p_j})\nonumber\\
 e^{(\lambda +\mu g_0)(p_{j+1}-p_j)}-1&\geq& g^a_1(p_{j+1}-p_j e^{(\lambda +\mu g_0)(p_{j+1}-p_j)})
 \label{ie}
\end{eqnarray} 
We also have
$g^b_1 = \frac{e^{(\lambda+\mu g_0)(p_j-p_{j+1})}-1}{p_j-p_{j+1} e^{(\lambda+\mu g_0)(p_j-p_{j+1})}}= \frac{e^{(\lambda+\mu g_0)(p_{j+1}-p_{j})}-1}{p_{j+1}-p_{j} e^{(\lambda+\mu g_0)(p_{j+1}-p_{j})}}$. 
Note that  $p_{j+1}>p_{j}$ implies $e^{(\lambda+\mu g_0)(p_{j+1}-p_{j})}-1>0$. Since $g^b_1 > 0$, we have $p_{j+1}-p_j e^{(\lambda +\mu g_0)(p_{j+1}-p_j)}>0$.  
Applying the above result to (\ref{ie}), we obtain,
$g^a_1\leq \frac{e^{(\lambda+\mu g_0)(p_{j+1}-p_{j})}-1}{p_{j+1}-p_{j} e^{(\lambda+\mu g_0)(p_{j+1}-p_{j})}}=g^b_1$
which is a contradiction to $g^a_1> g^b_1$.  Similarly, we  can also prove that if $p_j<p_{j+1}$,  we have $g^c_1\geq g^b_1$ which is a contradiction to $g^c_1< g^b_1$. Thus we must have $p_j>p_{j+1}$.
\subsection{Proof for Theorem \ref{l3} ii)} \label{appen3}
\indent \indent \indent \textit{Proof:}  From Lemma 2,  the boundary between any two adjacent regions  ${\cal{R}}_j$ and ${\cal{R}}_{j+1} $ is given by 
\begin{eqnarray}
g_1&=&\frac{e^{(\lambda+\mu g_0)(p_j-p_{j+1})}-1}{p_j-p_{j+1} e^{(\lambda+\mu g_0)(p_j-p_{j+1})}} = \frac{e^{(\lambda+\mu g_0) p_j}-e^{(\lambda+\mu g_0) p_{j+1}}}{p_j e^{(\lambda+\mu g_0) p_{j+1}}-p_{j+1} e^{(\lambda+\mu g_0) p_j}}\nonumber\\
&=&(\lambda+\mu g_0) \frac{e^{(\lambda+\mu g_0) p_\epsilon } (p_j-p_{j+1})}{p_j e^{(\lambda+\mu g_0) p_{j+1}}-p_{j+1} e^{(\lambda+\mu g_0) p_j}}
 > \lambda+\mu g_0
\end{eqnarray} 
where the last equality follows from the mean value theorem for some $p_\epsilon \in (p_{j+1}, p_j)$. The last inequality holds since we have $p_j e^{(\lambda+\mu g_0) p_\epsilon }> p_j e^{(\lambda+\mu g_0) p_{j+1} }$ and $-p_{j+1} e^{(\lambda+\mu g_0) p_\epsilon }> -p_{j+1} e^{(\lambda+\mu g_0) p_{j} }$.  By rearranging, we get $\frac{e^{(\lambda+\mu g_0) p_\epsilon } (p_j-p_{j+1})}{p_j e^{(\lambda+\mu g_0) p_{j+1}}-p_{j+1} e^{(\lambda+\mu g_0) p_j}}>1$. 
\subsection{Proof of Theorem \ref{l3} iii)} \label{appen4}
\indent \indent \indent \textit{Proof:}  Given  a fixed channel partitioning scheme, the optimal quantized power for ${\cal R}_j$ is obtained as $p_j=\max (p_j^*,0), \forall j$, where $p_j^*$ is determined by solving the equation $E[\frac{g_1}{1+g_1p_j}-(\lambda+\mu g_0)| {\cal{R}}_j]=0$. We can see that if $E[g_1| {\cal{R}}_j] \leq E[(\lambda+\mu g_0)| {\cal{R}}_j]$,  then to satisfy the equation,  $p_j^* <0$, implying  $p_j=\max (p_j^*,0)=0$. On the other hand,
if $E[g_1| {\cal{R}}_j] > E[(\lambda+\mu g_0)| {\cal{R}}_j]$, $p_j^*$  has to be strictly positive in order to satisfy the optimality 
equation, implying  $\max (p_j^*,0)=p_j^*$.  We know from Lemma \ref{l3} ii) that all boundaries between any two adjacent regions have a lower bound given by $g_1 > \lambda+\mu g_0$, i.e. 
for any given $(g_0,g_1)$ belonging to any of the first $L-1$ regions,    
$g_1  >\lambda+\mu g_0$. 
Thus for the first $L-1$ regions, 
$ E[g_1| {\cal{R}}_j] Pr\{ {\cal{R}}_j\} >
 E[(\lambda+\mu g_0)| {\cal{R}}_j]Pr\{ {\cal{R}}_j\}$ 
Therefore the optimal quantized power in the first $L-1$ regions is strictly positive. This cannot be said however for $p_L$ as for ${\cal R}_L$, 
we cannot guarantee $g_1 > \lambda+\mu g_0$ for any given $(g_0, g_1)$ pair in that region. It is thus possible to have $p_L$ to be zero. The next result shows under what circumstances 
one can have $p_L$ to be exactly $0$.
\subsection{Proof for Theorem \ref{l3} iv)} \label{appen5}
\indent \indent \indent \textit{Proof:} 
1) We know from Theorem \ref{l3} iii) that  we always have $E[\frac{g_1}{1+g_1p_j}-(\lambda+\mu g_0)| {\cal{R}}_j]=0, j=1,\dots,L-1,$ and for the region ${\cal{R}}_L$, this equation may not satisfied when $p_L=0$.  Let us assume that $p_L > 0$. Then we have
$\sum^L_{j=1} E[\lambda+\mu g_0| {\cal{R}}_j]Pr\{ {\cal{R}}_j\}=\sum^L_{j=1} E[\frac{g_1}{1+g_1p_j}| {\cal{R}}_j]Pr\{ {\cal{R}}_j\}$, implying
$\lambda+\mu=\sum^L_{j=1}E[\frac{g_1}{1+g_1p_j}| {\cal{R}}_j]Pr\{ {\cal{R}}_j\} <\sum^L_{j=1}E[{g_1}| {\cal{R}}_j]Pr\{ {\cal{R}}_j\}=1$ 
since $\sum_{j=1}^L E[g_i|{\cal{R}}_j]Pr\{ {\cal{R}}_j\} = E[g_i] =1$, for $i=0,1$. 
Hence if $\lambda+\mu\geq 1$, we must have $p_L=0$.\\
\indent From the optimality equation, one can write $p_i = \frac{E[\frac{g_1p_i}{1+g_1p_i} | {\cal R}_i]}{\lambda + \mu E[g_0 |{\cal R}_i]}$ when $p_i > 0$, it is obvious that $p_i < \frac{1}{\lambda + \mu E[g_0 |{\cal R}_i]}, \; i=1,2, \ldots, L-1$.  Since $p_L \geq 0$, this is also  true for region ${\cal{R}}_{L}$. Therefore when $\mu\not=0$,
$\mu Q_{avg} = \mu \sum_{i=1}^L p_i E[g_0|{\cal R}_i]Pr({\cal R}_i) < \sum_{i=1}^L \frac{\mu E[g_0|{\cal R}_i]}{\lambda + \mu E[g_0 |{\cal R}_i]} Pr({\cal R}_i) <\sum_{i=1}^L  Pr({\cal R}_i) = 1$.  Similarly, if $\lambda\not=0$, $ \lambda P_{avg}<1$. Thus $\mu > 1$ implies $Q_{av} < 1$ and $\lambda> 1$ implies $P_{av} < 1$.\\
\indent 2) Next, we will show that no matter what $\lambda, \mu$ is, $p_L$ must be zero for a 
sufficiently large $L$ and $\lim_{L\rightarrow \infty} p_{L-1}= 0$ . 
\begin{enumerate}
\item [(1)] First, we will prove that as $L\rightarrow \infty$, the  boundary between $ {\cal{R}}_{L-1}$ and $ {\cal{R}}_L$
approaches its limiting boundary $g_1=\frac{\lambda+\mu g_0}{1-(\lambda+\mu g_0) \delta^*}$, where $\delta^*=\lim_{L\rightarrow \infty} p_L$.
Given $p_1>\dots>p_L \geq 0$, it is clear that the sequence $\{p_i\},\; i=1,2, \ldots, L$ is a monotonically decreasing sequence bounded below, therefore it must converge to its greatest-lower bound $\delta^*$ ($\delta^*=\lim_{L\rightarrow \infty} p_L\geq 0$ ) as $L \rightarrow \infty$.
%(any bounded monotonically decreasing sequence is convergent and the limit is its greatest-lower bound ).
Therefore, it can be easily shown that 
for an arbitrarily small $\epsilon > 0$, we always can find  a sufficiently large $L$ such that $p_{L-1} - p_L < \epsilon$.
Thus,  as $L\rightarrow \infty$, $(p_{L-1}-p_L)\rightarrow 0$. Using this result, we can show that the  boundary between $ {\cal{R}}_{L-1}$ and $ {\cal{R}}_L$
approaches the limiting boundary $g_1=\frac{\lambda+\mu g_0}{1-(\lambda+\mu g_0) \delta^*}$ (or $\lambda+\mu g_0=\frac{g_1}{1+g_1\delta^*}$) as $L \rightarrow \infty$, (since this boundary can be written as $\lambda+\mu g_0=\frac{\log(\frac{1+g_1p_{L-1}}{1+g_1p_L})}{p_{L-1}-p_L}$, and $\lim_{L\rightarrow \infty}( \lim_{(p_{L-1}-p_L)\rightarrow 0}\frac{\log(\frac{1+g_1p_{L-1}}{1+g_1p_L})}{p_{L-1}-p_L})=\lim_{L\rightarrow \infty} \frac{g_1}{1+g_1p_L}=\frac{g_1}{1+g_1\delta^*}$).
\item [(2)]  Suppose there exists a ($\lambda, \mu$) such that $p_L > 0$ for any arbitrarily large  $L$ (implying $ \delta^*> 0$). Thus for any $L$, $p_L$ satisfies  $E[\frac{g_1}{1+g_1p_L}-(\lambda+\mu g_0)| {\cal{R}}_L]=0$. 
From (1), we have as $L \rightarrow \infty$, the boundary between $ {\cal{R}}_{L-1}$ and $ {\cal{R}}_L$ approaches its limit
 $\lambda+\mu g_0=\frac{g_1}{1+g_1\delta^*}$. Note that for a finite value of $L$, the region ${\cal R}_L$ can be divided into two parts ${\cal R}_{L1}$ and 
 ${\cal R}_{L2}$ where ${\cal R}_{L1}$ corresponds to $\frac{\log(\frac{1+g_1p_{L-1}}{1+g_1p_L})}{p_{L-1}-p_L} \leq \lambda+\mu g_0 < \frac{g_1}{1+g_1\delta^*}$ and 
 ${\cal R}_{L2}$ corresponds to 
 $ \frac{g_1}{1+g_1\delta^*} \leq \lambda+\mu g_0 < \infty$.  As $L$ becomes arbitrarily large, the region ${\cal R}_{L1}$ becomes vanishingly small, and 
 one obtains $E(\lambda+\mu g_0|{\cal R}_L)>E(\frac{g_1}{1+g_1\delta^*}|{\cal R}_L)\geq E(\frac{g_1}{1+g_1p_L}|{\cal R}_L)$ for a
sufficiently large $L$, which is a contradiction to the KKT optimality condition for $p_L > 0$. Hence no matter what $\lambda, \mu$ is, $p_L$ must be zero for a 
sufficiently large $L$. And $\delta^*=\lim_{L\rightarrow \infty} p_L=0$. 
\item [(3)]  $\delta^*=0$ implies the boundary between $ {\cal{R}}_{L-1}$ and $ {\cal{R}}_L$  approaches $g_1=\lambda+\mu g_0$ as $L \rightarrow \infty$, and since as $L\rightarrow \infty$, $(p_{L-1}-p_L) \rightarrow 0$ and $p_L= 0$, we have $\lim_{L\rightarrow \infty} p_{L-1}= 0$.
\end{enumerate}

\end{appendix}

\bibliographystyle{IEEEtran}

\begin{thebibliography}{1}
\providecommand{\url}[1]{#1}
\csname url@samestyle\endcsname
\providecommand{\newblock}{\relax}
\providecommand{\bibinfo}[2]{#2}
\providecommand{\BIBentrySTDinterwordspacing}{\spaceskip=0pt\relax}
\providecommand{\BIBentryALTinterwordstretchfactor}{4}
\providecommand{\BIBentryALTinterwordspacing}{\spaceskip=\fontdimen2\font plus
\BIBentryALTinterwordstretchfactor\fontdimen3\font minus
  \fontdimen4\font\relax}
\providecommand{\BIBforeignlanguage}[2]{{%
\expandafter\ifx\csname l@#1\endcsname\relax
\typeout{** WARNING: IEEEtran.bst: No hyphenation pattern has been}%
\typeout{** loaded for the language `#1'. Using the pattern for}%
\typeout{** the default language instead.}%
\else
\language=\csname l@#1\endcsname
\fi
#2}}
\providecommand{\BIBdecl}{\relax}
\BIBdecl
\bibitem{Ghasemi07}
A.~Ghasemi and E.~S.~Sousa, ``Fundamental limits of
spectrum-sharing in fading environments,'' \emph{IEEE Trans.
Wireless Commun.}, vol.~6, no.~2, pp.~649-658, Feb.~2007.
%\bibitem{Haykin05}
%S.~Haykin, ``Cognitive radio: brain-empowered 
%wireless communications,'' \emph{IEEE Journal on Selected Areas in Commun.}, vol.~23, no.~2, pp.~201-220, Feb.~2005.
\bibitem{Mitola99}
J.~Mitola III, ``Cognitive radio for flexible mobile multimedia communications,'' \emph{IEEE Int. Workshop on Mobile Multimedia Commun. (MoMuC) }, San Diego, CA, USA, Nov.~1999, pp.~3-10.
\bibitem{Goldsmith09}
A.~Goldsmith, S.A.~Jafar, I.~Maric, and S.~Srinivasa, ``Breaking spectrum gridlock with cognitive radios: an information theoretic perspective,'' \emph{Proceedings of the IEEE}, vol.~97, no.~5, pp.~894-914, May~2009.

\bibitem{peha09}
J.M.~Peha, ``Sharing Spectrum Through Spectrum Policy Reform and Cognitive Radio," {\em Proceedings of the IEEE}, vol.~97, no.~4, pp.~708--719, 
April 2009.
\bibitem{Zhang09}
R.~Zhang, ``On peak versus average interference power constraints
for protecting primary users in cognitive radio networks,''
\emph{IEEE Trans. Wireless Commun.}, vol.~8, no.~4, pp.~2112-2120,
April~2009.
\bibitem{Kang09}
X.~Kang, Y.~Liang, A.~Nallanathan, H.K.~Garg and R.~Zhang,
``Optimal power allocation for fading channels in cognitive radio
networks: Ergodic capacity and outage capacity,'' \emph{IEEE
Trans. Wireless Commun.}, vol.~8, no.~2, pp.~940-950, Feb.~2009.
\bibitem{Gastpar04}
M.~Gastpar, ``On capacity under received-signal constraints,'' \emph{42nd Annual Allerton Conf. on Commun., Control and Comp.}, Monticello, IL, USA,
Sept. 29 - Oct. 1, 2004.
%\bibitem{cogproofs}
%''http://sites.google.com/site/zhyfutopia/Home/cogproofs.pdf''
\bibitem{Linder99}
T.~Linder, and R.~Zamir, ``High-resolution source coding for non-difference distortion measures: the rate-distortion function,'' \emph{IEEE
Trans. Information Theory}, vol.~45, no.~2, pp.~533-547.
Mar.~1999.
\bibitem{Musavian09}
L. Musavian and S.~Aissa, ``Capacity and power allocation for spectrum-sharing communications in fading channels,''
\emph{IEEE Trans. Wireless Commun.}, vol.~8, no.~1, pp.~148-156, Jan.~2009.
\bibitem{Musavian092}
L. Musavian and S.~Aissa, ``Fundamental capacity limits of cognitive radio in fading environments with imperfect channel information,''
\emph{IEEE Trans. Commun.}, vol.~57, no.~11, pp.~3472-3480, Nov.~2009.

\bibitem{marques_giannakis09} 
A.G.~Marques, X.~Wang and G.B.~Giannakis, ``Dynamic Resource Management for Cognitive Radios Using Limited-Rate Feedback," 
{\em IEEE Transactions on Signal Processing}, vol. 57, no.~9, pp.~3651--3666, September 2009.
 
\bibitem{suraweera09}
H.A.~Suraweera, P.J.~Smith and M.~Shafi, ``Capacity Limits and Performance Analysis of Cognitive Radio With Imperfect Channel Knowledge," 
{\em IEEE Transactions on Vehicular Technology}, accepted for publication, 2010.  
%\bibitem{Marques09}
%A. G.~Marques, X.~Wang and G. B.~Giannakis, ``Dynamic resource management for cognitive radios using limited-rate feedback,''
%\emph{IEEE Trans. Signal Processing}, vol.~57, no.~9, pp.~3651-3666, Sep.~2009.
\bibitem{Linde80}
Y.~Linde, A.~Buzo, and R.~Gray, ``An Algorithm for Vector
Quantizer Design,'' \emph{IEEE Trans. Commun.}, vol.~28, no.~1,
pp.~84-95, Jan.~1980.
\bibitem{Gersho92}
A.~Gersho, and R.~Gray, ``Vector quantization and signal
compression,'' \emph{Kluwer Academic Publishers}, 1992.
%\bibitem{Roh06}
%J.C.~Roh and B.D.~Rao, ``Transmit beamforming in multiple-antenna systems with finite rate feedback: a VQ-based approach,'' \emph{IEEE
%Trans. Information Theory}, vol.~52, no.~3, pp.~1101-1112,
%Mar.~2006.
%\bibitem{Koyuncu08}
%E.~Koyuncu, Y.~Jing, and H.~Jafarkhani, ``Distributed beamforming in wireless relay networks with quantized feedback,'' \emph{IEEE Journal on Selected Areas in Commun.}, vol.~26, no.~8, pp.~1429-1439, Oct.~2008.
\bibitem{Sabin86}
M.~Sabin, and R.~Gray, ``Global convergence and empirical
consistency of the generalized Lloyd algorithm,'' \emph{IEEE
Trans. Information Theory}, vol.~32, no.~2, pp.~148-155,
Mar.~1986.
\bibitem{Zhang08}
L.~Zhang, Y.~Xin and Y.~Liang,``Optimal power allocation for
multiple access channels in cognitive radio networks,'' in
\emph{Proc. IEEE Vehicular Technology Conference (VTC Spring
2008)}, Singapore, 11-14 May 2008, pp.~1550-2252.
\bibitem{RZhang09}
R.~Zhang, S.~Cui and Y.~Liang ``On ergodic sum capacity of fading cognitive multiple-access and
broadcast channels,'' \emph{IEEE
Trans. Information Theory}, vol.~55, no.~11, pp.~5161-5178,
Nov.~2009.
\bibitem{Ekbatani09}
 S.~Ekbatani, F.~Etemadi and H.~Jafarkhani, ``Throughput maximization over slowly fading channels using quantized and erroneous feedback," {\emph IEEE Transactions on Communications}, 
vol. 57, no.~9, pp.~2528-2533, Sep. 2009.

\bibitem{Zeger90}
K.~Zeger and A.~Gersho, ``Pseudo-Gray coding ," {\emph IEEE Transactions on Communications}, 
vol. 38, no.~12, pp.~2147-2158, Dec. 1990.

\bibitem{KRZhang09}
K.~Huang and R.~Zhang ``Cooperative feedback for multi-antenna cognitive radio networks ,'' \emph{arXiv:0911.2952}, 
Submitted on 16 Nov 2009.

\bibitem{zhang_sigprocmag}
R.~Zhang, Y-C.~Liang and S.~Cui, ``Dynamic Resource Allocation in Cognitive Radio Networks," {\em IEEE Signal Processing Magazine}, 
vol. 27, no.~3, pp.~102-114, March 2010.

\end{thebibliography}

%\hfill

\end{document}